\documentclass[%
 reprint,
 amsmath,amssymb,
 aps,
]{revtex4-2}
\usepackage{float}
\usepackage{graphicx}% Include figure files
\usepackage{dcolumn}% Align table columns on decimal point
\usepackage{bm}% bold math
\usepackage{mathtools}
\usepackage{tikz}
\usepackage{caption}
\usepackage{subcaption}

\begin{document}

\title{Demixing in binary mixtures with differential diffusivity at high density}
\author{Erin McCarthy}
\affiliation{Department of Physics and BioInspired Institute,
Syracuse University, Syracuse, New York, 13244}
\author{Ojan Damavandi}
\affiliation{Department of Physics and BioInspired Institute,
Syracuse University, Syracuse, New York, 13244}
\author{Raj Kumar Manna}
 \email{rajkmphys@gmail.com}
\affiliation{Department of Physics and BioInspired Institute,
Syracuse University, Syracuse, New York, 13244}
\author{M. Lisa Manning}
 \email{mmanning@syr.edu}
\affiliation{Department of Physics and BioInspired Institute,
Syracuse University, Syracuse, New York, 13244}
\date{\today}

\begin{abstract}
Spontaneous phase separation, or demixing, is important in biological phenomena such as cell sorting. In particle-based models, an open question is whether differences in diffusivity can drive such demixing. While differential-diffusivity-induced phase separation occurs in mixtures with a packing fraction up to $0.7$~\cite{weber_2016}, here we investigate whether demixing persists at even higher densities relevant for cells. For particle packing fractions between $0.7$ and $1.0$ the system demixes, but at packing fractions above unity the system remains mixed, exposing re-entrant behavior in the phase diagram. We also find that a confluent Voronoi model for tissues does not phase separate, consistent with the highest-density particle-based simulations.
\end{abstract}

\maketitle

%Introduction
Spontaneous sorting is a common emergent behavior in particle packings composed of different species. While surface tension in fluid-like mixtures composed of molecules with different adhesion is a canonical mechanism that drives phase separation, there are many particulate systems that phase separate due to other mechanisms.

One example is sorting in vibrated granular materials, where sorting occurs when particles of the same size differ in density, as well as when particles of the same density differ in size~\cite{rivas_segregation_2011, rivas_sudden_2011}. Sorting also occurs in thermal systems; for example, an entropy-based depletion force causes phase separation for large colloidal particles suspended in a solution of smaller particles ~\cite{mao_depletion_1995,biben_depletion_1996,louis_effective_2002,dijkstra_evidence_1994,marenduzzo_depletion_2006,asakura_interaction_1958}. 

In biological materials, cell sorting and phase separation contribute to compartmentalization and patterning of groups of cells, especially during development~\cite{heller_tissue_2015, steinberg_reconstruction_nodate, harris_is_1976, brodland_differential_2002, akam_making_1989, chamberlain_notochord-derived_2008, christian_xwnt-8_1991, krens_cell_2011}. At the subcellular level, the ability of particles to undergo robust, spontaneous segregation allows for organization within the cell membrane~\cite{rayermann_hallmarks_2017, cornell_tuning_2018, banjade_phase_2014, miller_membrane_2018}, protein compartmentalization~\cite{li_phase_2012, patel_liquid--solid_2015, bieler_whole-embryo_2011}, and the formation of non-membrane bound organelles~\cite{hyman_anthony_a_controlling_nodate, su_phase_2016, brangwynne_germline_2009, hyman_active_2011}. 

Given the complexity of biological materials, there are multiple physical mechanisms that could be driving these sorting behaviors, including standard mechanisms of surface tension and differential adhesion. However, even in systems without adhesion, materials that are active can also spontaneously phase separate. One such phenomenon is called Motility Induced Phase Separation (MIPS)~\cite{cates_motility-induced_2015, marchetti_hydrodynamics_2013}, which occurs when persistently moving particles drive a feedback loop between velocity and density that leads to a droplet-forming instability. A related mechanism can also help drive phase separation in mixtures of self-propelled particles at intermediate and high densities~\cite{yang2014aggregation}.

Importantly, Weber and coworkers~\cite{weber_2016} have demonstrated that the time invariance introduced by persistent self-propulsion is not required for phase separation; ordinary diffusion is enough, provided that diffusivity is different for two different species. Specifically, they studied mixtures of Brownian repulsive particles where the two different particle species differed only in their diffusivity. While such a system is active because one of the species is not equilibrated with the thermal bath, particle motion is not persistent. In such a system, the authors find that at low and intermediate densities (up to a packing fraction of $0.7$) differential diffusivity drives phase separation via nucleation and coarsening into cold droplets surrounded by a hot gas. 

Similar results have been observed in a model of active and passive dumbbells, up to a packing fraction of 1.0~\cite{nayana_soft_2021}. In addition, recent analytical work has shown that in the dilute limit, hard spheres with differential diffusivity exhibit a positive effective surface tension at the interface and undergo binodal and spinodal decomposition~\cite{ilker_2020}. This result is only approximate when a solid phase exists, and third-order corrections change the phase diagram significantly, suggesting that particle packings at intermediate and high densities could exhibit emergent behaviors not captured by the analytic theory. 

Understanding the role of differential diffusivity at higher densities is important in real biological systems.  Recent work has demonstrated that enzymes in the presence of their substrates behave as active particles that are not persistent and have an enhanced diffusivity~\cite{xu_direct_2019}, and these processes occur in a dense intercellular environment. In addition, some cell types exhibit differential diffusivity compared to their neighbors inside aggregates \emph{in vitro}~\cite{ZevGartner}. Given these results, an open question is whether differential diffusivity can drive segregation at high densities, relevant for the crowded environment inside the cell, inside multicellular aggregates, or in dense active colloidal suspensions.

It is known that homogeneous packings of active particles change their behavior dramatically at high densities, as glassy dynamics emerge. Previous work has shown that homogeneous packings of soft disks with self-propulsion reach a new, glassy state at high densities~\cite{henkes_2011}. Such packings exhibit complex dynamics, including avalanches~\cite{morse_direct_2021} and intermittent plasticity~\cite{mandal_extreme_2020}. 

Therefore, the goal of this Letter is to study whether the glassy dynamics that emerge at high densities alter or destroy the differential-diffusivity-induced phase separation seen in low- and intermediate- density particulate systems. To address this question, we numerically investigate a repulsive two-species particulate system at intermediate and high densities and quantify the phase separation over a wide range of model parameters. For completeness, we also simulate and analyze dynamics in a model for confluent systems, where there are no gaps or overlaps between cells/particles.

\begin{figure}[h]
    \centering
    \includegraphics[width=\linewidth]{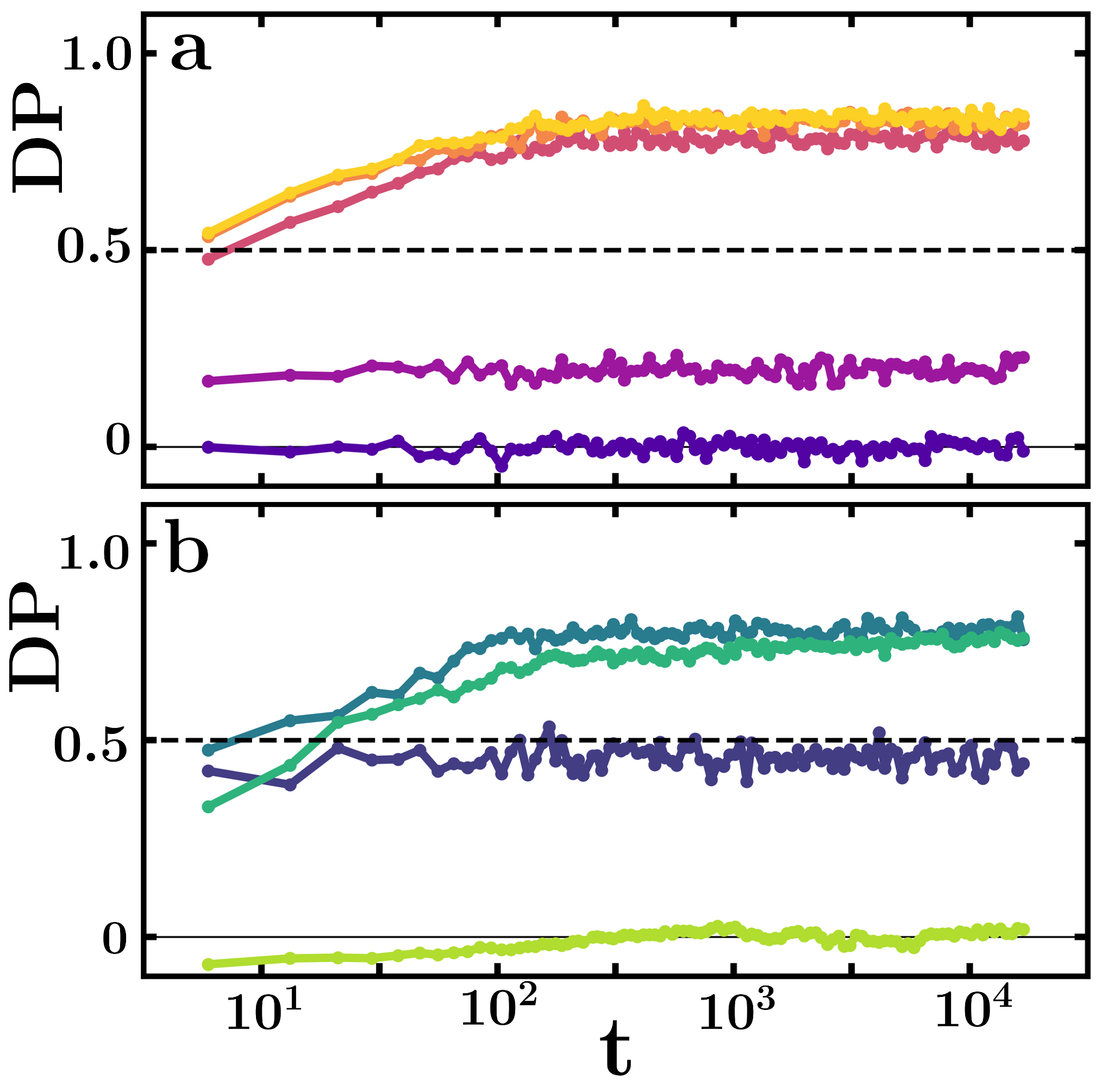}
    \caption{\label{fig:1}
    {\bf Demixing in binary mixtures of particles with diffusion constants $D_{\text{cold}}$ and $D_{\text{hot}}$ for various values of the ratio $D=D_{\text{cold}}/D_{\text{hot}}$ and packing fraction, $\phi$.} a) The time evolution of demixing parameter (DP) with fixed $\phi = 0.6$ for varying $D= 1.0, 0.1, 0.01, 0.001,$ and $0.0001$, shown in purple, magenta, pink, orange, and yellow, respectively. (b) Time evolution with $D = 0.01$ for varying $\phi = 0.3, 0.6, 0.9$, and $1.2$, shown in light green, green, teal, and indigo, showing non-monotonic behavior in $\phi$.
    }
    \end{figure}

\begin{figure*}[!ht]
    \centering
    \includegraphics[width=0.95\linewidth]{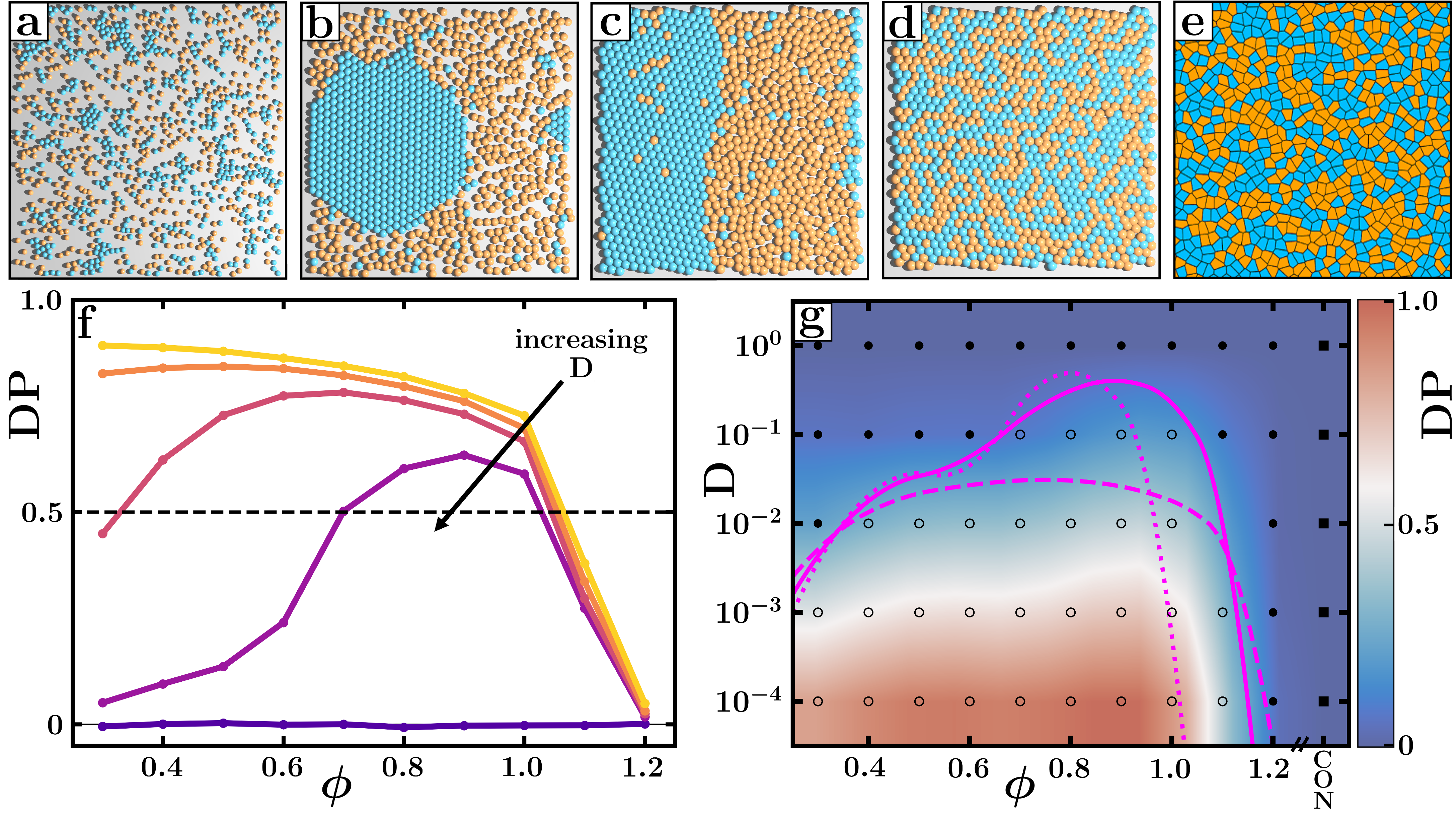}
    \caption{\label{fig:2}
    {\bf Re-entrant behavior of a binary mixture of particles with differential diffusivity.}  (a-d) Steady-state configurations of cold (blue spheres) and hot (orange spheres) particles  at $\phi = 0.3$, $0.6$, $0.9$, and $1.2$ respectively with $D=0.01$ and  $D_{hot}=5.0$. (e) The long-time behavior of the Voronoi system at the same value of D. (f) The average demixing parameter (DP) is plotted against $\phi$ for different values of $D$, using the same colormap described in Fig.~\ref{fig:1}(a). (g) Phase diagram depicting no large-scale demixing (open circles) and large-scale demixing (closed circles) in $\phi$ and $D$ parameter space. The Voronoi systems are marked by solid squares.  The background is colored by interpolated values of  DP at $D_{hot}=5.0$. The dotted, solid, and dashed lines separate the areas of the phase of large-scale demixing with no demixing respectively for $D_{hot}= 2.0$,  $D_{hot}=$ 5.0, and  $D_{hot}=$ 10.0. (See SM for phase diagrams for $D_{hot}= 2.0, 10$.)
    }
    \end{figure*}
%Methods
We consider a binary particle mixture where the particle types differ only in their diffusivity. The diffusion constants for $``\text{cold}"$ and $``\text{hot}"$ particles are $D_{\text{cold}}$ and $D_{\text{hot}}$ respectively where $D_{\text{cold}} \le D_{\text{hot}}$  and the ratio of their diffusion constants is defined as $D=D_{\text{cold}}/D_{\text{hot}}$. The particles interact with soft repulsive Hertzian potential, $E=k(1-r_{ij}/2R)^{5/2}$ for $r_{ij}<2R$, here $r_{ij}=|\bm{r}_{ij}|=|\bm{r}_i-\bm{r}_j|$ is the distance between two particles and $R$ and $k$ are the radius and stiffness of the particles. The particle dynamics is governed by over-damped Langevin equations of motion (see SM for details). We initialize a 50:50 mixture of cold and hot particles distributed uniformly in a box of length $L$ and evolve the system with time. We investigate the system with different $D$ and packing fraction, $\phi=N\pi R^2/L^2$, where $N$ is the total number of particles in the system. Unless otherwise noted, we simulate $N=1000$ particles with $D_{hot} = 5.0$, as we have verified (SM Fig S2) that value of $D_{hot}$ is sufficiently high to drive diffusive particle behavior at the highest packing fractions studied. Results are reported in natural units of length $R$ and time $\tau= R^{2}/D_{hot}$. To complement our results at higher packing fractions, we also study a confluent Voronoi model~\cite{bi_motility-driven_2016} where there are no gaps between particles (see SM for details).

%Results
We first confirm that for sufficiently small values of $D$ our model shows large-scale demixing at low packing fractions, where cold particles form a large cluster, surrounded by gas-like hot particles  as seen in Ref.~\cite{weber_2016}. We quantify the amount of demixing using the demixing parameter, $\text{DP}=\langle DP_i \rangle =\langle 2(N_{s}/N_{t}-0.5)\rangle$, where $N_s$  and $N_t$ are respectively the number of homotypic neighbors and the total number of neighbors of particle $i$. For $D<1.0$, DP evolves to reach a steady-state plateau indicating the  demixing configurations is favorable in the system, as shown in Fig.~\ref{fig:1}(a). As discussed in other work~\cite{sahu_small-scale_2020}, it is possible for the demixing parameter to reach a small, but  non-zero steady state value if a system partially demixes on small scales.  This microscopic demixing is observed for $D=0.1$ at $\phi=0.6$ (see Fig. \ref{fig:1}a).

Next, we study how the demixing of particles changes with the packing fraction, $\phi$ for a fixed $D$. For small values of $D$, there is no large-scale demixing at very low packing fraction (Fig.~\ref{fig:1}b and Fig.~\ref{fig:2}a), but systems exhibit macroscopic large-scale demixing at intermediate values of $\phi$ (Fig.~\ref{fig:2}b-c). However, large-scale demixing becomes less favorable as the packing fraction increases and there is less free space between particles. As shown in Fig.~\ref{fig:1}b and Fig.~\ref{fig:2}d, the large-scale demixing of particles does not occur at very high packing fractions ($\phi \sim 1.1-1.2$).  We confirm the results of no demixing at high densities with a Voronoi model of a binary mixture of particles (see SM for details) where there is no gap between particles (Fig.~\ref{fig:2}e).  This validates our hypothesis that free space between particles is necessary for large-scale demixed configurations.

How does the demixing of the particles vary across $\phi$ and $D$ parameter space? To answer this, we plot the steady state demixing parameter, averaged over $10$ ensembles as a function of $\phi$ for different values of $D$ in Fig.~\ref{fig:2}f. For all values of $D$,  differential diffusivity of particles cannot drive large-scale demixing at the highest packing fraction $\phi=1.2$. 
For certain values of $D$, the system does not demix at both very low and very high values of $\phi$ and exhibits large-scale demixing at intermediate values of $\phi$. Therefore we observe a re-entrant behavior in the $\phi$ direction.

For the system sizes considered here, large-scale demixing is sharply defined by a DP value greater than 0.5 (see SM Fig S1(a)). Using this threshold DP value, we construct a phase diagram (Fig.~\ref{fig:2}g) in $\phi$ and $D$ parameter space that marks the regions of large-scale demixing states (open circles). The solid pink line separates the region of large-scale demixing with no-large scale demixing at $D_{\text{hot}}=5.0$.  We also investigate the impact of overall system temperature on the phase diagram by varying the diffusivity of hot particles, $D_{\text{hot}}$. The qualitative features of the phase diagram remain the same for different values of $D_{\text{hot}}$ (dashed and dotted pink lines in Fig.~\ref{fig:2}g, see SM for full phase diagrams). Additionally, the results of the confluent Voronoi model (where there is no free space between particles) are indicated by the label ‘CON’ that shows no demixing at any value of $D$. 

\begin{figure}[]
    \includegraphics[width=85mm]{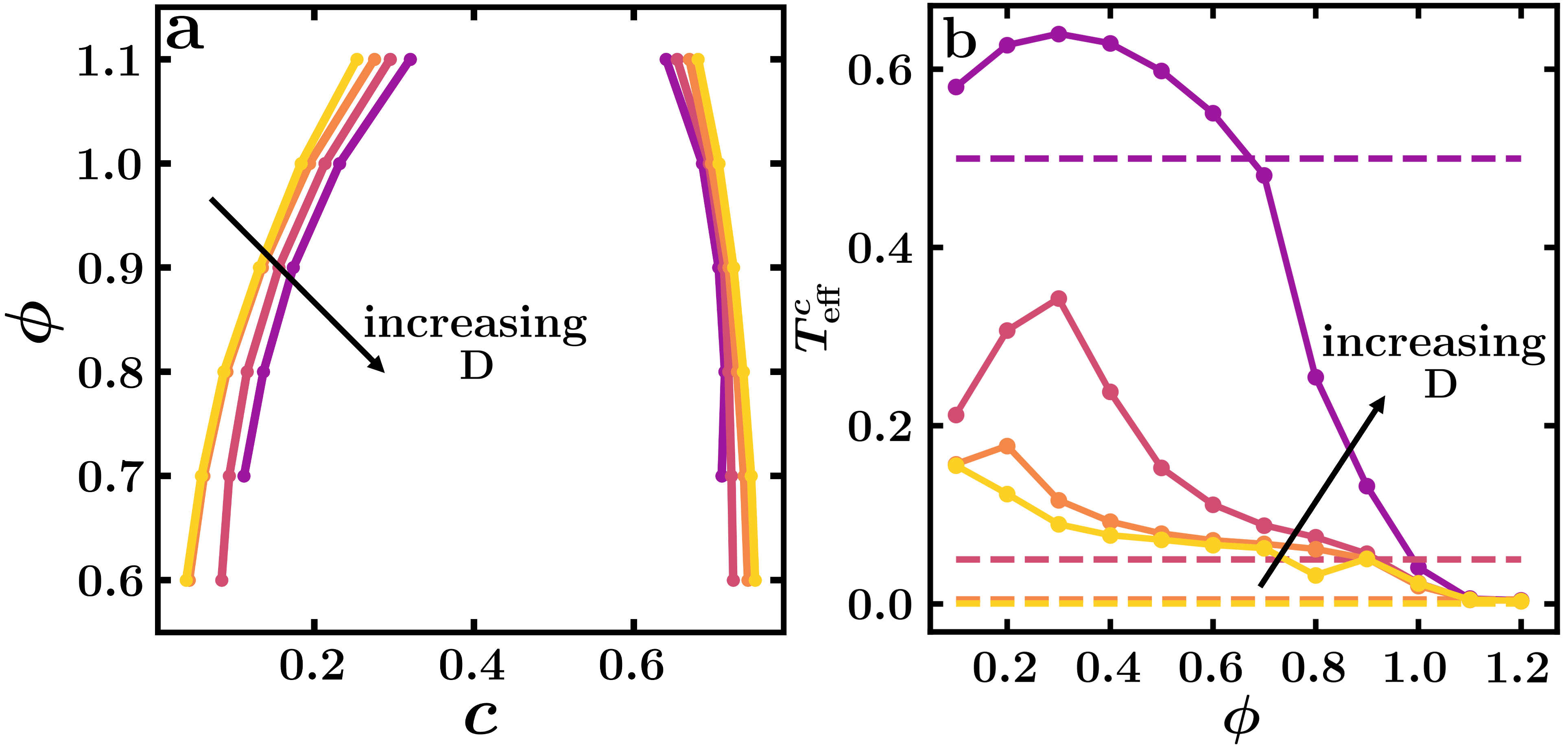}
    \caption{\label{fig:3} {\bf Intrinsic properties of mixed and demixed phases.} (a) Binodal lines from the concentration $c$ of cold particles inside ($c_{in}$, RHS) and outside ($c_{out}$, LHS) of a condensed cluster for $D = 0.1, 0.01, 0.001,$ and $0.0001$. Colormap is the same as in Fig.~\ref{fig:1}a.  (b) Effective temperature $T^c_{eff}$ extracted from the mean-squared displacement of cold particles as a function of $\phi$ for the same values of $D$ shown in (a). Dotted lines indicate the temperature derived from the input diffusivity of cold particles. 
 }
    \end{figure}

To further understand the demixing behavior of the particle mixtures, for demixed states with $D \le 0.1$ we construct a binodal based upon the concentration $c$ of cold particles inside ($c_{in}$) and outside ($c_{out}$) of the condensed cluster, where the concentration is extracted from a cumulative density distribution (see SM for details). As shown in Fig.~\ref{fig:3}a,  the two concentration profiles for ($c_{in}$) and ($c_{out}$) would meet at the center above $\phi=1.1$, indicating that no demixing occurs and there is one concentration throughout the packing. Therefore, for $D \le 0.1$, there is a critical point in the system above $\phi=1.1$ (at a system size of $N=1000$) beyond which demixing is no longer favorable.

We expect that this re-entrant phase transition is related to the entropy of hot and cold particles, but directly computing entropy production in very dense systems is challenging. Instead, we were able to quantify an effective temperature~\cite{smrek_small_2017} of cold particles, $T_{\text{eff}}^{\text{c}}$, extracted from the mean-squared displacement (see SM for details), which appears to be relevant for the phase behavior across the entire range of packing fractions studied. At low $\phi$, effective temperatures are higher than the input temperature due to interactions with the hot particles, while at high $\phi$ the effective temperatures are lower due to caging effects. The packing fraction at which  $T_{\text{eff}}^{c}$ becomes maximum defines the onset of the demixed state (Fig.~\ref{fig:3}b). The effective temperature also drops to a very small value at the re-entrant transition where the system re-mixes and where neighbor cages significantly restrict particle motion.

%Conclusions

In summary, at very high densities differential diffusivity does not drive demixing in particle-based systems. Mixtures of soft particles with different diffusivities exhibit re-entrant behavior, where phase separation is not observed at low densities nor very high densities. Similarly, a Voronoi model for a confluent system where there are no gaps between particles never demixes due to differential diffusivity. These results indicate that free space between particles is necessary for diffusivity-based demixing. 

%Discussion
Given that we only see the re-entrant behavior for packing fractions above unity, one might wonder whether such high packing fractions are relevant or reasonable. They are certainly relevant; best-fit particle-based models for dense biological tissues are often in a regime with a packing fraction greater than unity~\cite{petridou2021rigidity,ranft2010fluidization}. In addition, the mechanical and dynamical behavior of soft-sphere particle packings is reasonably robust up to packing fractions of at least $1.2-1.3$, before next-nearest neighbor interactions become important and lead to new minimum-energy structures~\cite{ellenbroek2013unusual}.

Previous analytic work~\cite{ilker_2020} on systems with differential diffusivity predicts demixing as the density increases and does not predict re-entrant behavior, but the existing theory does not include many-body corrections. It would therefore be interesting to investigate whether including third- or higher-order interactions, even perturbatively, would suggest a suppression of demixing.

In addition, our observation that free volume is required for sorting suggests a quasi-thermodynamic picture for the observed re-entrant behavior. At intermediate densities, sorting of the low-diffusivity particles into a compact phase gives high-diffusivity particles access to more configurations, which may increase the total entropy. At high enough densities, however, the lack of free volume prevents the high diffusivity particles from accessing additional configurations. 

A challenge to testing this hypothesis directly is that it is difficult to define analogues of thermodynamic quantities in a system where particles at two different temperatures do not equilibrate. While the kinetic pressure is a valid observable in molecular dynamics simulations and swim pressure is useful for thermodynamic-like describes of phase separation in self-propelled particle systems~\cite{takatori_towards_2015}, in this Brownian system we were unable to define a pressure-like variable that could capture phase behavior. We also found that standard methods for calculating entropy production in glassy materials~\cite{smrek_small_2017} failed at the highest densities, though were were able to compute a related effective temperature intensive variable that is informative. Future work could focus on identifying other metrics for the phase transition.

A potentially related observation is that at higher densities, even in regions of phase space that do not macroscopically demix, we observe local demixing over a length scale of a few particle or cell diameters.  Such ``micro-demixing" is quite robust for all values of $D$. A similar phenomenon was recently observed in a confluent cell model composed of two different cell types that differed in their preferred cell shape~\cite{sahu_small-scale_2020}. In that case, the origin was demonstrated to be purely kinetic; differential energy barriers to cell rearrangement exist for cells of one type to enter an island of another type, even though the states themselves are energetically equivalent. It would be interesting to consider whether differential diffusivity similarly generates asymmetric rates of rearrangement at high densities.

%We have conjectured that free volume may be crucial in facilitating demixing due to differences in diffusivity, but further work should be done to verify this hypothesis analytically, perhaps using an order parameter, such as a modified pressure, to predict the location of the critical point. We did attempt to find a thermodynamic explanation of our results using standard definitions of interaction pressure, kinetic pressure, and swim pressure. However, our results were inconclusive, as it is non-trivial to characterize the active pressure in a system without formal directed motion such as self-propulsion. At certain systems near the critical point, we observed microdemixing, where particles do not fully phase-separate but rather undergo small-scale patterning. Capturing the mechanism driving this emergent behavior may also be important in developing a complete understanding of this system.

This work has immediate implications for biological systems. Most obviously, different cell types in confluent or nearly confluent biological tissues have been observed to have different diffusivities. An open question is whether such differential diffusivity could be responsible for cell sorting seen in such experiments. Our work demonstrates the answer is unequivocally no -- researchers must search for other differences between the cell types, such as heterotypic interfacial tensions~\cite{canty_sorting_2017, sussman_soft_2018}, to drive sorting.

Another potential application is the collective behavior of enzymes in dense cellular environments. It has recently been shown that enzymes in the presence of their substrate diffuse as Brownian particles at a higher effective temperature than their surroundings~\cite{xu_direct_2019}. Given the extremely crowded environment inside cells~\cite{dix_crowding_2008, andre_liquidliquid_2020}, as well as the increased concentration of active enzymes in liquid-liquid phase separation~\cite{peeples_mechanistic_2021} it may be interesting to investigate whether differential diffusivity could help segregate enzymes at intermediate densities, and whether the re-mixing we observe at high enough densities could also occur for collections of enzymes inside cells. 

%Enzymes are known to exhibit differences in diffusivity \cite{xu_direct_2019} so this result could provide further understanding of the underlying mechanics of the crucial enzymatic activity within the cell. However, when considering behavior on the cellular level within tissues, environments are often very crowded. Therefore, our results suggest that diffusivity based demixing can be ruled out as the mechanism of many cases of cellular sorting

Taken together, our results highlight that active sorting behaviors at high densities do not extrapolate in a simple way from observations at intermediate densities, and suggest that investigations of the high-density limit in other active matter systems may also uncover non-trivial emergent behavior.
%that we should explicitly investigate the high density limit in other active processes.
   % \item The re-entrant behavior observed at high densities serves as an example that sorting is a complex behavior that highly depends on the density of a system

%Given the importance of spontaneous sorting to many biological processes, our results emphasize the need to investigate sorting mechanisms at high densities in order to fully understand their relevance to biological processes.
{ \bf Acknowledgements} The authors acknowledge support from NSF-DMR-1951921 (EM and MLM) NIH R01HD099031 (RKM and MLM), and Simons Foundation 454947 (OKD and MLM). 

\section{Supplementary Material}

\renewcommand{\thefigure}{S\arabic{figure}}
\setcounter{figure}{0}
\setcounter{section}{0}
\renewcommand\thesection{\Alph{section}}
%\section{\label{sec:Methods}Extended Methods}
\section{\label{sec:PBM}Description of Particle-Based Simulations}
We model a binary mixture soft spheres of uniform size and shape in a periodic box. The two cell types differ only by their diffusivity, where $N_{hot}+N_{cold}=N=1000$. Hot particles have a diffusion constant of $D_{hot}$, and cold particles have a diffusion constant of $D_{cold}$. Cell-cell interactions are dictated by a hertzian energy potential.

\begin{eqnarray}
    E= \begin{cases*}
        k\delta^{\frac{5}{2}} & \text{if $r_{ij}<2R$}\\
        0 & \text{otherwise}
    \end{cases*};
    \\
    \delta = 1-\frac{r_{ij}}{2R} \text{ (if } r_{ij} < 2R).
\end{eqnarray}

where $k$ represents particle stiffness,  $R$ is the radius of a single particle, and $\mathbf{r}_{ij}=\mathbf{r}_{i}-\mathbf{r}_{j}$. The force on a particle $\mathbf{F}_{ij}$ is found by 
\begin{eqnarray}
    \mathbf{F}_{ij}=\frac{\partial{E}_{ij}}{\partial\mathbf{r}_{ij}},
\end{eqnarray}
such that $\mathbf{F}_{ij}=0$ when particle overlap is equal to 0, and $\mathbf{F}_{ij}$ becomes non-zero when $\delta>0$ because $|\mathbf{r}_{ij}|<2R$. The dynamics of the packing are determined by the sum of these short range interactions, and a translational noise term ($\bm{\eta}_{i}$), described in detail below: 

\begin{eqnarray}
    \frac{d\mathbf{r}_{i}}{dt}=\mu\sum_{j\neq i}\mathbf{F}_{ij}+D_{type}\bm{\eta}_{i},
\\
    D_{type}=D_{hot}\:\mathrm{or}\:D_{cold}.
\end{eqnarray}
Here $\bm{\eta}_{i}$ is a numerical approximation to Gaussian white noise with an average of $0$ and a standard deviation of $1$, where $\langle \eta _{i\alpha}(t)\eta _{j\beta}(t')\rangle = \delta_{ij}\delta_{\alpha\beta}\delta(t-t')$, which means that the noise on different particles $i$ and $j$ are uncorrelated with one another and uncorrelated in time, so that the noise is memoryless. The strength of the noise is set by the diffusivity $D_{type}$. In practice, this means that in our numerical simulation at each time step of unit $dt$, the position is incremented by a Gaussian random step drawn from a distribution with standard deviation $\sqrt{2dtD_{type}}$.

The particle stiffness $k$ is set to a value of $500$ and the time step $dt$ is set to a value of $0.001$ to ensure that the numerical integration scheme is stable and does not depend on our choice of $dt$. $\mu$ is the mobility or inverse drag coefficient, which is set to $1$. The natural time unit, $\tau$, is equal to $R^{2}/D_{hot}$. The value of $D_{hot}$, which determines the overall level of activity in the system, is set to a value of $5.0$, unless otherwise noted, such that the system remains fluid-like even at high densities and the strength of the activity does not become too large for our time step. However, we also perform a sweep in $D_{hot}$ to investigate changes to the phase-diagram when the overall amount of energy varies. The other two parameters we vary are the packing fraction, represented by $\phi$, and the parameter $D$, representing the ratio between the diffusivities of the two particle types:

\begin{eqnarray}
    D=\frac{D_{hot}}{D_{cold}}.
\end{eqnarray}

\begin{figure*}
\includegraphics[width=.7\linewidth]{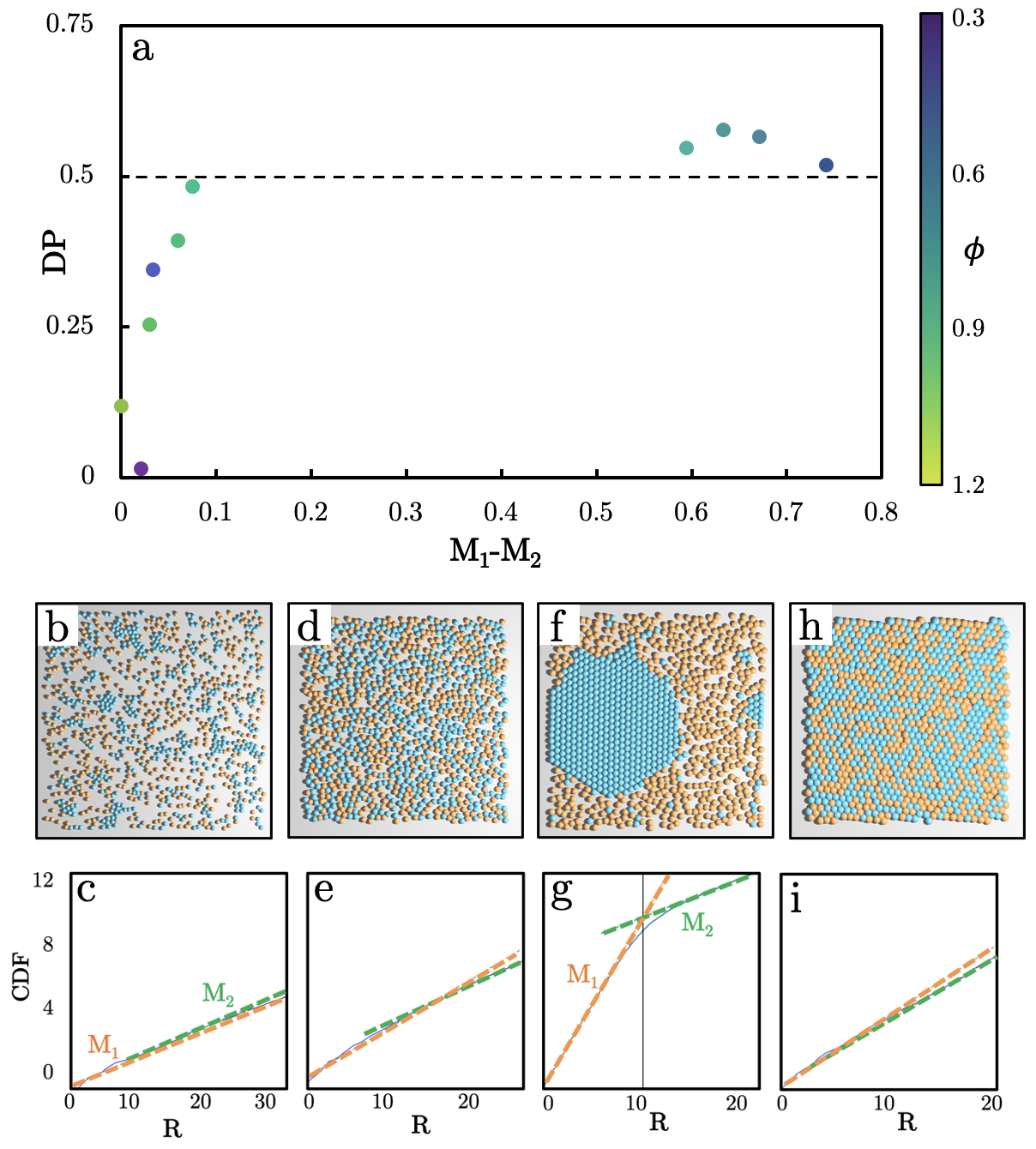}
\caption{\label{fig:OP} {\bf Quantifying droplet formation and concentration.}
(a) The long-time plateau value of DP against the difference in slopes extracted from the CDF for different values of $\phi$. Each point is colored by the $\phi$ value of that system, as shown in the colorbar on the right. All systems are for $D = 10^{-3}$. (b) Snapshot at a late time point from the system at $\phi=0.3$, and (c) CDF of this system. The slope from the inner section of the box is plotted in orange ($M_{1}$), and the slope from the outer section of the box is plotted in green ($M_{2}$). Panels (d-i) show the same snapshots and quantities for $\phi= 0.6, 0.9 and 1.2$ from left to right.}
\end{figure*}

\section{\label{sec:Voronoi}Description of Voronoi Simulations}
To further understand activity-based phase-separation at high densities, we model a confluent system where again two particle types differ only by their diffusion constants. We use a Voronoi model~\cite{bi_motility-driven_2016} with parameters identical to those of the particle-based model, aside from the energy functional of the packing. Particle dynamics are determined by minimizing the energy function $E = E(\mathbf{r}_{i})$
\begin{eqnarray}
    E=\sum_{i=j}^{N}[K_{A}(A(\mathbf{r_{i}})-A_{0})^{2}+K_{P}(P(\mathbf{r_{i}})-P_{0})^{2}].
\end{eqnarray}
The area term represents volume incompressability, and the perimeter term results from contractility of the cytoskeleton, as well as a cell-membrane tension due to cell-cell adhesion. $p_{0}$ is a non-dimensionalized shape index equal to $P_{0}/\sqrt{A_{0}}$. Contrasting the entirely localized force interactions of the particle-based model, forces in the confluent system are non-local and non-additive, such that $\mathbf{F}_{i}=-\mathbf{\nabla}_{i}E$. Dynamics of the system are determined by
\begin{eqnarray}
    \frac{d\mathbf{r}_{i}}{dt}=\mu\mathbf{F}_{i}++D_{type}\bm{\eta}_{i},
\end{eqnarray}d
where $\mu$, the noise term and $D_{type}$ are the same as described for the particulate system. 

\section{\label{sec:AQ}Analysis of simulation data}

The Demixing Parameter (DP) is used to quantify the amount of demixing in a packing. In the limit of large system sizes, when cells are completely sorted by type, we expect DP to give a value of 1; in a completely mixed system, we expect a value of 0.
\begin{eqnarray}
    DP=\langle DP_{i}\rangle =\langle 2(\frac{N_{s}}{N_{t}}-\frac{1}{2})\rangle,
\end{eqnarray}
where $N_{s}$ is the number of homotypic neighbors of particle $i$, and $N_{t}$ is the total number of neighbors of particle $i$. When DP gives a value of 0.5 or above, we consider the system to have undergone large-scale demixing. We determine the DP threshold beyond which the system has formed a droplet or stripe by comparing DP values to the difference in slopes found using the Cumulative Density Function, described below; these values can be seen plotted in Fig.~\ref{fig:OP}a. \\

We use a Cumulative Density Function (CDF) to characterize the change in density between the solid and gas phases, as well as to determine the radius of a formed droplet. This function sums the number of particles within small circular bins size dr moving out from the center of mass of the cold particles. In a uniform system, the numerical density, $n(r)$ is expected to be a constant value. In our system, we expect it to be one constant value inside the cluster and another outside. The concentration of particles, $c(r)$, has the same expectations, as it is simply the numerical denisty scaled with the particle area. In our system, the area of a particle is equal to 1, therefore, simply counting up the particles and taking the derivative can give us a concentration directly. The CDF is calculated as follows:
\begin{eqnarray}
    CDF(r)=\sum{\frac{N_{bin}}{A_{bin}}}=\int_{0}^{maxR}c(r) dr.
\end{eqnarray}

\begin{figure*}[]
\includegraphics[width=\linewidth]{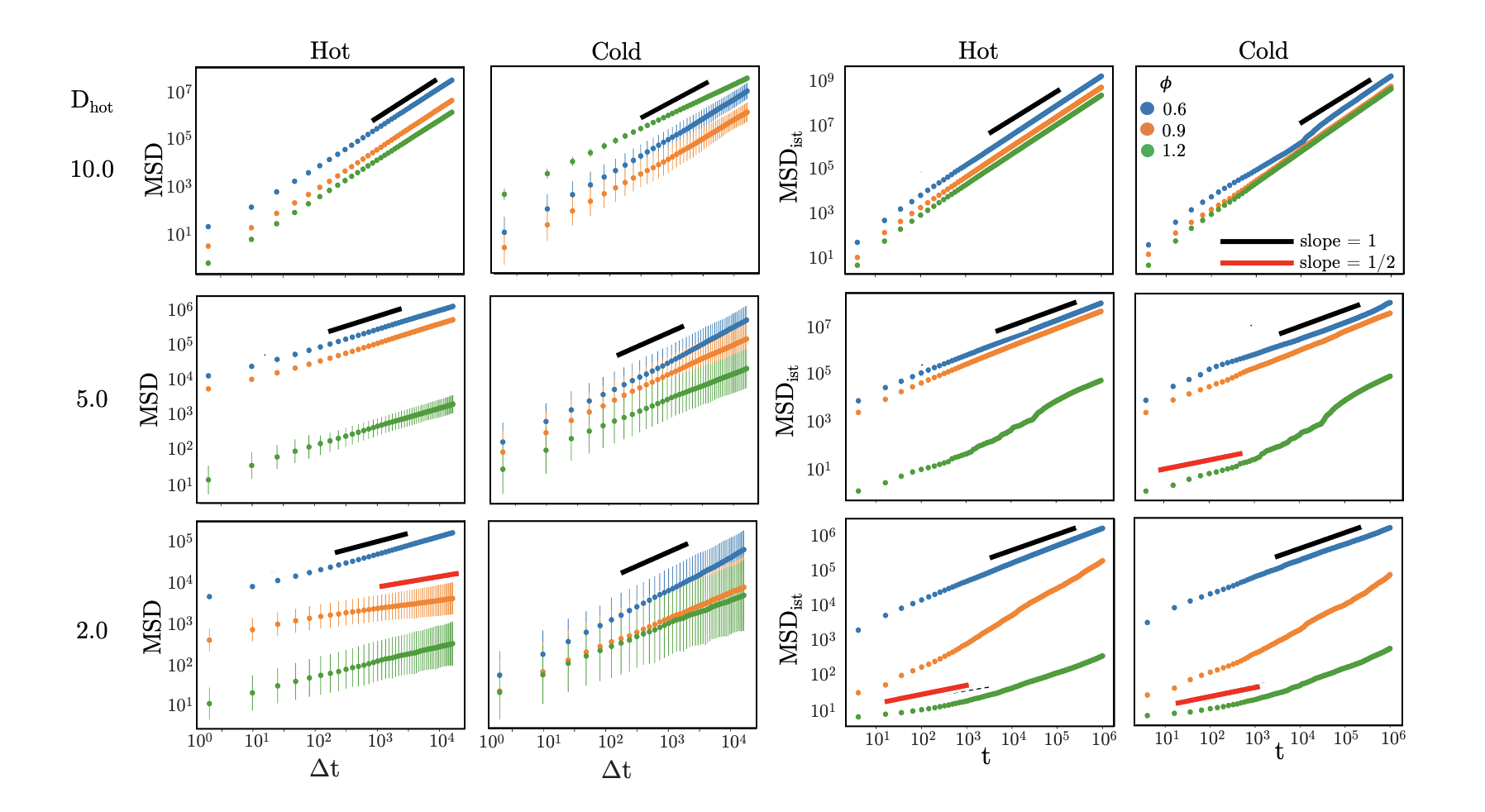}
\caption{\label{fig:MSD} 
Logarithmically-scaled mean squared displacement (MSD) and mean squared distance (MSD$_{\text{ist}}$) of only the hot particles, and only the cold particles of systems at various $D_{hot}$ values ($2.0, 5.0$ and $10.0$). Each plot displays data from three values of $\phi$, 0.6, 0.9 and 1.2, shown in blue, orange and green, respectively. The MSD plots have error bars scaled by the standard deviation over ten runs. On each plot, slope lines are shown to guide the eye, with a slope of $1$ represented by a black line and a slope of $1/2$ represented by a red line.}
\end{figure*}

Where $N_{bin}$ is the number of particles in a bin, and $A_{bin}$ is the area of the bin. Therefore, after calculating this function from the center of mass to the edge of the box, taking a derivative will output the concentration at that part of the box (slope=M$_{1}$ for the inner section, and slope=M$_2$ for the outer section). From this, we can find $c_{in}$, the concentration inside the cluster, and $c_{out}$, the concentration outside the cluster, in systems where a droplet forms. These values can then be used to construct a binodal, as shown in Figure 3 in the main text. Additionally, by finding the value of $r$ at which the slope of the CDF changes, we can find the radius of a formed droplet.  If there is a significant difference between $M_{1}$ and $M_{2}$, this indicates that a droplet has formed. As shown in Fig. \ref{fig:OP}(a), for our system size, $M_{1}-M_{2}$ has a sharp bimodal distribution, where it is less than $0.1$ if a droplet has not formed and is greater than $0.5$ if a droplet is formed. Moreover, a value of DP $= 0.5$ provides a sharp threshold for this jump, and therefore we use DP to quantify the phase transition. Fig.~\ref{fig:OP}(b-i) show snapshots and CDF plots for different values of $\phi$. \\

\begin{figure*}[]
\includegraphics[width=.8\linewidth]{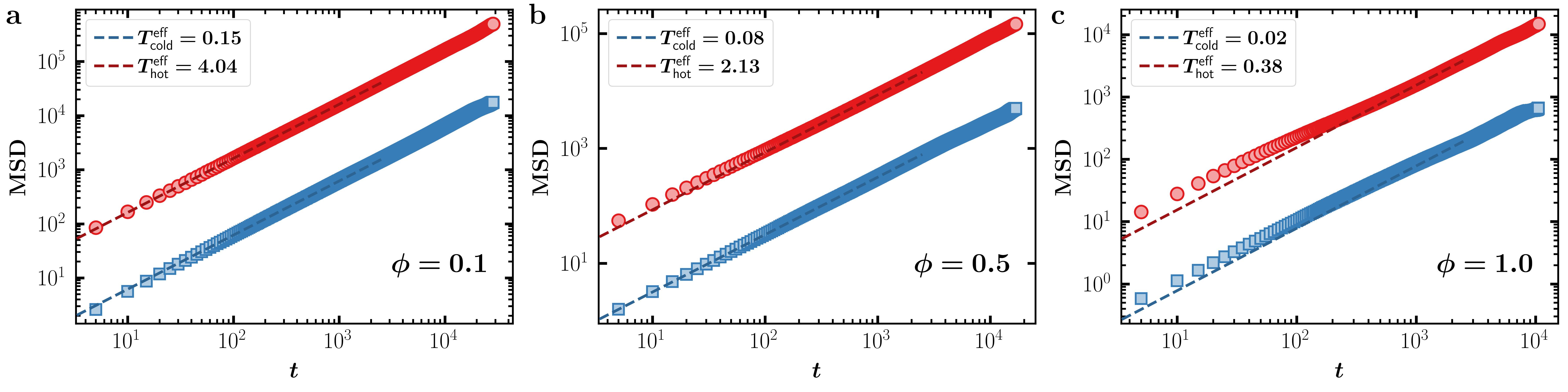}
\caption{\label{fig:MSD2} 
Logarithmically-scaled mean squared displacement (MSD) for hot and cold particles for packing fraction 0.1 (a) and 0.5 (b) and 1.0 (c). The input temperatures of cold and hot particles are respectively 0.005 and 5.0 ($D_{\text{hot}}=5$ and $D=0.001$).  }
\end{figure*}
Mean squared displacement(MSD) is another metric used to quantify the behavior of the system. MSD is a measure of how far a particle travels on average in some given amount of time, called $\Delta T$. Therefore we find
\begin{eqnarray}
\delta r_{t}^{2}(\Delta T) = (r_{x}(t + \Delta T) - r_{x}(t))^{2} +
(r_{y}(t + \Delta T) - r_{y}(t))^{2},
\end{eqnarray}

where $r_{x}$ is the particle's position in the x direction and $r_{y}$ is the y position. This value is then averaged over many t values throughout the simulation such that,
\begin{eqnarray}
MSD(\Delta T) = \frac{1}{N_{t}}\sum_{t} \delta r_{t}^{2}(\Delta T).
\end{eqnarray}
MSD is a useful measure of how much particles are moving around in a global sense. Particularly, we use MSD to ensure that, as we approach high densities and particles become caged by their neighbors, the system can still explore different configurations. For particles whose motion is diffusive, MSD scales linearly with time. Therefore, on a log-log plot of MSD vs time, diffusive particles will give a line with a slope of 1. For our system, if MSD has a slope close to unity, that indicates fluid-like behavior at long times. A slope of $1/2$ indicates that particle motion is sub-diffusive, meaning particles must be, to some degree, caged by their neighbors. 

For active systems that become so crowded that particles cannot change neighbors at all, MSD is no longer an effective tool because there can be a mode where the entire box is translating so that particles move without changing neighbors. In these cases, we use Mean Squared Distance (MSD$_{\text{ist}}$), a similar measurement that removes those translations by considering the distances between a particle and the particle that was its closest neighbor at $t=0$. We have the same expectations of the slope when this quantity is plotted against time logarithmically.

Log-log plots of both MSD and MSD$_{\text{ist}}$ are shown in Fig~\ref{fig:MSD}. When $D_{hot}=$ 10.0, both MSD and MSD$_{\text{ist}}$ have a slope near unity for both hot and cold particles. When $D_{hot}=$ 5.0, MSD has a slope near 1 for both hot and cold particles. However, for the cold particles at a $\phi$ value of $1.2$, MSD$_{\text{ist}}$ does have a slope near $1/2$ at early time points, although it eventually reaches a slope near 1. This indicates that at very high density, cold particles are trapped by their neighbors on a short time scale, but given enough time they are able to change neighbors and display diffusive behavior. When $D_{hot}=$ 2.0, MSD$_{\text{ist}}$ has a slope near 1/2 for both hot and cold particles at $\phi$ = 1.2.
This indicates that the system becomes arrested and sold-like at high density, when $D_{hot}=$ 2.0. Therefore, a $D_{hot}$ value of at least $5.0$ is needed for the system to remain fluid-like on long timescales. \\

The effective temperatures $T_{\text{hot}}^{\text{eff}}$ and $T_{\text{cold}}^{\text{eff}}$ respectively for hot and cold particles are computed by measuring long-time diffusivity from mean-squared displacement.

\begin{eqnarray}
MSD_{\text{type}}(t) = 4  D_{\text{type}}^{\text{eff}} t,
\end{eqnarray}
where $D_{\text{type}}^{\text{eff}}=\mu k_B T_{\text{type}}^{\text{eff}}$ is the effective diffusivity for either hot or cold particles. The effective temperatures obtained for three packing fractions for $D_{\text{hot}}=5$ and $D=0.001$ are shown in Fig.~\ref{fig:MSD2}.

\bibliography{references.bib}

%apsrev4-2.bst 2019-01-14 (MD) hand-edited version of apsrev4-1.bst
%Control: key (0)
%Control: author (8) initials jnrlst
%Control: editor formatted (1) identically to author
%Control: production of article title (0) allowed
%Control: page (0) single
%Control: year (1) truncated
%Control: production of eprint (0) enabled
\begin{thebibliography}{50}%
\makeatletter
\providecommand \@ifxundefined [1]{%
 \@ifx{#1\undefined}
}%
\providecommand \@ifnum [1]{%
 \ifnum #1\expandafter \@firstoftwo
 \else \expandafter \@secondoftwo
 \fi
}%
\providecommand \@ifx [1]{%
 \ifx #1\expandafter \@firstoftwo
 \else \expandafter \@secondoftwo
 \fi
}%
\providecommand \natexlab [1]{#1}%
\providecommand \enquote  [1]{``#1''}%
\providecommand \bibnamefont  [1]{#1}%
\providecommand \bibfnamefont [1]{#1}%
\providecommand \citenamefont [1]{#1}%
\providecommand \href@noop [0]{\@secondoftwo}%
\providecommand \href [0]{\begingroup \@sanitize@url \@href}%
\providecommand \@href[1]{\@@startlink{#1}\@@href}%
\providecommand \@@href[1]{\endgroup#1\@@endlink}%
\providecommand \@sanitize@url [0]{\catcode `\\12\catcode `\$12\catcode
  `\&12\catcode `\#12\catcode `\^12\catcode `\_12\catcode `\%12\relax}%
\providecommand \@@startlink[1]{}%
\providecommand \@@endlink[0]{}%
\providecommand \url  [0]{\begingroup\@sanitize@url \@url }%
\providecommand \@url [1]{\endgroup\@href {#1}{\urlprefix }}%
\providecommand \urlprefix  [0]{URL }%
\providecommand \Eprint [0]{\href }%
\providecommand \doibase [0]{https://doi.org/}%
\providecommand \selectlanguage [0]{\@gobble}%
\providecommand \bibinfo  [0]{\@secondoftwo}%
\providecommand \bibfield  [0]{\@secondoftwo}%
\providecommand \translation [1]{[#1]}%
\providecommand \BibitemOpen [0]{}%
\providecommand \bibitemStop [0]{}%
\providecommand \bibitemNoStop [0]{.\EOS\space}%
\providecommand \EOS [0]{\spacefactor3000\relax}%
\providecommand \BibitemShut  [1]{\csname bibitem#1\endcsname}%
\let\auto@bib@innerbib\@empty
%</preamble>
\bibitem [{\citenamefont {Weber}\ \emph {et~al.}(2016)\citenamefont {Weber},
  \citenamefont {Weber},\ and\ \citenamefont {Frey}}]{weber_2016}%
  \BibitemOpen
  \bibfield  {author} {\bibinfo {author} {\bibfnamefont {S.~N.}\ \bibnamefont
  {Weber}}, \bibinfo {author} {\bibfnamefont {C.~A.}\ \bibnamefont {Weber}},\
  and\ \bibinfo {author} {\bibfnamefont {E.}~\bibnamefont {Frey}},\ }\bibfield
  {title} {\bibinfo {title} {Binary mixtures of particles with different
  diffusivities demix},\ }\bibfield  {journal} {\bibinfo  {journal} {Physical
  Review Letters}\ }\textbf {\bibinfo {volume} {116}},\ \href
  {https://doi.org/10.1103/physrevlett.116.058301}
  {10.1103/physrevlett.116.058301} (\bibinfo {year} {2016})\BibitemShut
  {NoStop}%
\bibitem [{\citenamefont {Rivas}\ \emph
  {et~al.}(2011{\natexlab{a}})\citenamefont {Rivas}, \citenamefont {Cordero},
  \citenamefont {Risso},\ and\ \citenamefont {Soto}}]{rivas_segregation_2011}%
  \BibitemOpen
  \bibfield  {author} {\bibinfo {author} {\bibfnamefont {N.}~\bibnamefont
  {Rivas}}, \bibinfo {author} {\bibfnamefont {P.}~\bibnamefont {Cordero}},
  \bibinfo {author} {\bibfnamefont {D.}~\bibnamefont {Risso}},\ and\ \bibinfo
  {author} {\bibfnamefont {R.}~\bibnamefont {Soto}},\ }\bibfield  {title}
  {\bibinfo {title} {Segregation in quasi-two-dimensional granular systems},\
  }\href {https://doi.org/10.1088/1367-2630/13/5/055018} {\bibfield  {journal}
  {\bibinfo  {journal} {New J. Phys.}\ }\textbf {\bibinfo {volume} {13}},\
  \bibinfo {pages} {055018} (\bibinfo {year} {2011}{\natexlab{a}})},\ \bibinfo
  {note} {publisher: IOP Publishing}\BibitemShut {NoStop}%
\bibitem [{\citenamefont {Rivas}\ \emph
  {et~al.}(2011{\natexlab{b}})\citenamefont {Rivas}, \citenamefont {Ponce},
  \citenamefont {Gallet}, \citenamefont {Risso}, \citenamefont {Soto},
  \citenamefont {Cordero},\ and\ \citenamefont {Mujica}}]{rivas_sudden_2011}%
  \BibitemOpen
  \bibfield  {author} {\bibinfo {author} {\bibfnamefont {N.}~\bibnamefont
  {Rivas}}, \bibinfo {author} {\bibfnamefont {S.}~\bibnamefont {Ponce}},
  \bibinfo {author} {\bibfnamefont {B.}~\bibnamefont {Gallet}}, \bibinfo
  {author} {\bibfnamefont {D.}~\bibnamefont {Risso}}, \bibinfo {author}
  {\bibfnamefont {R.}~\bibnamefont {Soto}}, \bibinfo {author} {\bibfnamefont
  {P.}~\bibnamefont {Cordero}},\ and\ \bibinfo {author} {\bibfnamefont
  {N.}~\bibnamefont {Mujica}},\ }\bibfield  {title} {\bibinfo {title} {Sudden
  {Chain} {Energy} {Transfer} {Events} in {Vibrated} {Granular} {Media}},\
  }\href {https://doi.org/10.1103/PhysRevLett.106.088001} {\bibfield  {journal}
  {\bibinfo  {journal} {Phys. Rev. Lett.}\ }\textbf {\bibinfo {volume} {106}},\
  \bibinfo {pages} {088001} (\bibinfo {year} {2011}{\natexlab{b}})},\ \bibinfo
  {note} {publisher: American Physical Society}\BibitemShut {NoStop}%
\bibitem [{\citenamefont {Mao}\ \emph {et~al.}(1995)\citenamefont {Mao},
  \citenamefont {Cates},\ and\ \citenamefont
  {Lekkerkerker}}]{mao_depletion_1995}%
  \BibitemOpen
  \bibfield  {author} {\bibinfo {author} {\bibfnamefont {Y.}~\bibnamefont
  {Mao}}, \bibinfo {author} {\bibfnamefont {M.~E.}\ \bibnamefont {Cates}},\
  and\ \bibinfo {author} {\bibfnamefont {H.~N.~W.}\ \bibnamefont
  {Lekkerkerker}},\ }\bibfield  {title} {\bibinfo {title} {Depletion force in
  colloidal systems},\ }\href {https://doi.org/10.1016/0378-4371(95)00206-5}
  {\bibfield  {journal} {\bibinfo  {journal} {Physica A: Statistical Mechanics
  and its Applications}\ }\textbf {\bibinfo {volume} {222}},\ \bibinfo {pages}
  {10} (\bibinfo {year} {1995})}\BibitemShut {NoStop}%
\bibitem [{\citenamefont {Biben}\ \emph {et~al.}(1996)\citenamefont {Biben},
  \citenamefont {Bladon},\ and\ \citenamefont
  {Frenkel}}]{biben_depletion_1996}%
  \BibitemOpen
  \bibfield  {author} {\bibinfo {author} {\bibfnamefont {T.}~\bibnamefont
  {Biben}}, \bibinfo {author} {\bibfnamefont {P.}~\bibnamefont {Bladon}},\ and\
  \bibinfo {author} {\bibfnamefont {D.}~\bibnamefont {Frenkel}},\ }\bibfield
  {title} {\bibinfo {title} {Depletion effects in binary hard-sphere fluids},\
  }\href {https://doi.org/10.1088/0953-8984/8/50/008} {\bibfield  {journal}
  {\bibinfo  {journal} {J. Phys.: Condens. Matter}\ }\textbf {\bibinfo {volume}
  {8}},\ \bibinfo {pages} {10799} (\bibinfo {year} {1996})},\ \bibinfo {note}
  {publisher: IOP Publishing}\BibitemShut {NoStop}%
\bibitem [{\citenamefont {Louis}\ \emph {et~al.}(2002)\citenamefont {Louis},
  \citenamefont {Allahyarov}, \citenamefont {Löwen},\ and\ \citenamefont
  {Roth}}]{louis_effective_2002}%
  \BibitemOpen
  \bibfield  {author} {\bibinfo {author} {\bibfnamefont {A.~A.}\ \bibnamefont
  {Louis}}, \bibinfo {author} {\bibfnamefont {E.}~\bibnamefont {Allahyarov}},
  \bibinfo {author} {\bibfnamefont {H.}~\bibnamefont {Löwen}},\ and\ \bibinfo
  {author} {\bibfnamefont {R.}~\bibnamefont {Roth}},\ }\bibfield  {title}
  {\bibinfo {title} {Effective forces in colloidal mixtures: {From} depletion
  attraction to accumulation repulsion},\ }\href
  {https://doi.org/10.1103/PhysRevE.65.061407} {\bibfield  {journal} {\bibinfo
  {journal} {Phys. Rev. E}\ }\textbf {\bibinfo {volume} {65}},\ \bibinfo
  {pages} {061407} (\bibinfo {year} {2002})},\ \bibinfo {note} {publisher:
  American Physical Society}\BibitemShut {NoStop}%
\bibitem [{\citenamefont {Dijkstra}\ and\ \citenamefont
  {Frenkel}(1994)}]{dijkstra_evidence_1994}%
  \BibitemOpen
  \bibfield  {author} {\bibinfo {author} {\bibfnamefont {M.}~\bibnamefont
  {Dijkstra}}\ and\ \bibinfo {author} {\bibfnamefont {D.}~\bibnamefont
  {Frenkel}},\ }\bibfield  {title} {\bibinfo {title} {Evidence for
  entropy-driven demixing in hard-core fluids},\ }\href
  {https://doi.org/10.1103/PhysRevLett.72.298} {\bibfield  {journal} {\bibinfo
  {journal} {Phys. Rev. Lett.}\ }\textbf {\bibinfo {volume} {72}},\ \bibinfo
  {pages} {298} (\bibinfo {year} {1994})},\ \bibinfo {note} {publisher:
  American Physical Society}\BibitemShut {NoStop}%
\bibitem [{\citenamefont {Marenduzzo}\ \emph {et~al.}(2006)\citenamefont
  {Marenduzzo}, \citenamefont {Finan},\ and\ \citenamefont
  {Cook}}]{marenduzzo_depletion_2006}%
  \BibitemOpen
  \bibfield  {author} {\bibinfo {author} {\bibfnamefont {D.}~\bibnamefont
  {Marenduzzo}}, \bibinfo {author} {\bibfnamefont {K.}~\bibnamefont {Finan}},\
  and\ \bibinfo {author} {\bibfnamefont {P.~R.}\ \bibnamefont {Cook}},\
  }\bibfield  {title} {\bibinfo {title} {The depletion attraction: an
  underappreciated force driving cellular organization},\ }\href
  {https://doi.org/10.1083/jcb.200609066} {\bibfield  {journal} {\bibinfo
  {journal} {Journal of Cell Biology}\ }\textbf {\bibinfo {volume} {175}},\
  \bibinfo {pages} {681} (\bibinfo {year} {2006})}\BibitemShut {NoStop}%
\bibitem [{\citenamefont {Asakura}\ and\ \citenamefont
  {Oosawa}(1958)}]{asakura_interaction_1958}%
  \BibitemOpen
  \bibfield  {author} {\bibinfo {author} {\bibfnamefont {S.}~\bibnamefont
  {Asakura}}\ and\ \bibinfo {author} {\bibfnamefont {F.}~\bibnamefont
  {Oosawa}},\ }\bibfield  {title} {\bibinfo {title} {Interaction between
  particles suspended in solutions of macromolecules},\ }\href
  {https://doi.org/10.1002/pol.1958.1203312618} {\bibfield  {journal} {\bibinfo
   {journal} {Journal of Polymer Science}\ }\textbf {\bibinfo {volume} {33}},\
  \bibinfo {pages} {183} (\bibinfo {year} {1958})}\BibitemShut {NoStop}%
\bibitem [{\citenamefont {Heller}\ and\ \citenamefont
  {Fuchs}(2015)}]{heller_tissue_2015}%
  \BibitemOpen
  \bibfield  {author} {\bibinfo {author} {\bibfnamefont {E.}~\bibnamefont
  {Heller}}\ and\ \bibinfo {author} {\bibfnamefont {E.}~\bibnamefont {Fuchs}},\
  }\bibfield  {title} {\bibinfo {title} {Tissue patterning and cellular
  mechanics},\ }\href {https://doi.org/10.1083/jcb.201506106} {\bibfield
  {journal} {\bibinfo  {journal} {J Cell Biol}\ }\textbf {\bibinfo {volume}
  {211}},\ \bibinfo {pages} {219} (\bibinfo {year} {2015})}\BibitemShut
  {NoStop}%
\bibitem [{\citenamefont {Steinberg}()}]{steinberg_reconstruction_nodate}%
  \BibitemOpen
  \bibfield  {author} {\bibinfo {author} {\bibfnamefont {M.~S.}\ \bibnamefont
  {Steinberg}},\ }\href
  {https://www-science-org.libezproxy2.syr.edu/doi/10.1126/science.141.3579.401}
  {\bibinfo {title} {Reconstruction of {Tissues} by {Dissociated} {Cells}
  {\textbar} {Science}}}\BibitemShut {NoStop}%
\bibitem [{\citenamefont {Harris}(1976)}]{harris_is_1976}%
  \BibitemOpen
  \bibfield  {author} {\bibinfo {author} {\bibfnamefont {A.~K.}\ \bibnamefont
  {Harris}},\ }\bibfield  {title} {\bibinfo {title} {Is cell sorting caused by
  differences in the work of intercellular adhesion? {A} critique of the
  steinberg hypothesis},\ }\href {https://doi.org/10.1016/0022-5193(76)90019-9}
  {\bibfield  {journal} {\bibinfo  {journal} {Journal of Theoretical Biology}\
  }\textbf {\bibinfo {volume} {61}},\ \bibinfo {pages} {267} (\bibinfo {year}
  {1976})}\BibitemShut {NoStop}%
\bibitem [{\citenamefont {Brodland}(2002)}]{brodland_differential_2002}%
  \BibitemOpen
  \bibfield  {author} {\bibinfo {author} {\bibfnamefont {G.~W.}\ \bibnamefont
  {Brodland}},\ }\bibfield  {title} {\bibinfo {title} {The {Differential}
  {Interfacial} {Tension} {Hypothesis} ({DITH}): {A} {Comprehensive} {Theory}
  for the {Self}-{Rearrangement} of {Embryonic} {Cells} and {Tissues}},\ }\href
  {https://doi.org/10.1115/1.1449491} {\bibfield  {journal} {\bibinfo
  {journal} {Journal of Biomechanical Engineering}\ }\textbf {\bibinfo {volume}
  {124}},\ \bibinfo {pages} {188} (\bibinfo {year} {2002})}\BibitemShut
  {NoStop}%
\bibitem [{\citenamefont {Akam}(1989)}]{akam_making_1989}%
  \BibitemOpen
  \bibfield  {author} {\bibinfo {author} {\bibfnamefont {M.}~\bibnamefont
  {Akam}},\ }\bibfield  {title} {\bibinfo {title} {Making stripes
  inelegantly},\ }\href {https://doi.org/10.1038/341282a0} {\bibfield
  {journal} {\bibinfo  {journal} {Nature}\ }\textbf {\bibinfo {volume} {341}},\
  \bibinfo {pages} {282} (\bibinfo {year} {1989})},\ \bibinfo {note} {number:
  6240 Publisher: Nature Publishing Group}\BibitemShut {NoStop}%
\bibitem [{\citenamefont {Chamberlain}\ \emph {et~al.}(2008)\citenamefont
  {Chamberlain}, \citenamefont {Jeong}, \citenamefont {Guo}, \citenamefont
  {Allen},\ and\ \citenamefont {McMahon}}]{chamberlain_notochord-derived_2008}%
  \BibitemOpen
  \bibfield  {author} {\bibinfo {author} {\bibfnamefont {C.~E.}\ \bibnamefont
  {Chamberlain}}, \bibinfo {author} {\bibfnamefont {J.}~\bibnamefont {Jeong}},
  \bibinfo {author} {\bibfnamefont {C.}~\bibnamefont {Guo}}, \bibinfo {author}
  {\bibfnamefont {B.~L.}\ \bibnamefont {Allen}},\ and\ \bibinfo {author}
  {\bibfnamefont {A.~P.}\ \bibnamefont {McMahon}},\ }\bibfield  {title}
  {\bibinfo {title} {Notochord-derived {Shh} concentrates in close association
  with the apically positioned basal body in neural target cells and forms a
  dynamic gradient during neural patterning},\ }\href
  {https://doi.org/10.1242/dev.013086} {\bibfield  {journal} {\bibinfo
  {journal} {Development}\ }\textbf {\bibinfo {volume} {135}},\ \bibinfo
  {pages} {1097} (\bibinfo {year} {2008})}\BibitemShut {NoStop}%
\bibitem [{\citenamefont {Christian}\ \emph {et~al.}(1991)\citenamefont
  {Christian}, \citenamefont {McMahon}, \citenamefont {McMahon},\ and\
  \citenamefont {Moon}}]{christian_xwnt-8_1991}%
  \BibitemOpen
  \bibfield  {author} {\bibinfo {author} {\bibfnamefont {J.~L.}\ \bibnamefont
  {Christian}}, \bibinfo {author} {\bibfnamefont {J.~A.}\ \bibnamefont
  {McMahon}}, \bibinfo {author} {\bibfnamefont {A.~P.}\ \bibnamefont
  {McMahon}},\ and\ \bibinfo {author} {\bibfnamefont {R.~T.}\ \bibnamefont
  {Moon}},\ }\bibfield  {title} {\bibinfo {title} {Xwnt-8, a {Xenopus}
  {Wnt}-1/int-1-related gene responsive to mesoderm-inducing growth factors,
  may play a role in ventral mesodermal patterning during embryogenesis},\
  }\href {https://doi.org/10.1242/dev.111.4.1045} {\bibfield  {journal}
  {\bibinfo  {journal} {Development}\ }\textbf {\bibinfo {volume} {111}},\
  \bibinfo {pages} {1045} (\bibinfo {year} {1991})}\BibitemShut {NoStop}%
\bibitem [{\citenamefont {Krens}\ and\ \citenamefont
  {Heisenberg}(2011)}]{krens_cell_2011}%
  \BibitemOpen
  \bibfield  {author} {\bibinfo {author} {\bibfnamefont {S.~F.~G.}\
  \bibnamefont {Krens}}\ and\ \bibinfo {author} {\bibfnamefont {C.-P.}\
  \bibnamefont {Heisenberg}},\ }\bibfield  {title} {\bibinfo {title} {Cell
  sorting in development},\ }\href
  {https://doi.org/10.1016/B978-0-12-385065-2.00006-2} {\bibfield  {journal}
  {\bibinfo  {journal} {Curr Top Dev Biol}\ }\textbf {\bibinfo {volume} {95}},\
  \bibinfo {pages} {189} (\bibinfo {year} {2011})}\BibitemShut {NoStop}%
\bibitem [{\citenamefont {Rayermann}\ \emph {et~al.}(2017)\citenamefont
  {Rayermann}, \citenamefont {Rayermann}, \citenamefont {Cornell},
  \citenamefont {Merz},\ and\ \citenamefont
  {Keller}}]{rayermann_hallmarks_2017}%
  \BibitemOpen
  \bibfield  {author} {\bibinfo {author} {\bibfnamefont {S.~P.}\ \bibnamefont
  {Rayermann}}, \bibinfo {author} {\bibfnamefont {G.~E.}\ \bibnamefont
  {Rayermann}}, \bibinfo {author} {\bibfnamefont {C.~E.}\ \bibnamefont
  {Cornell}}, \bibinfo {author} {\bibfnamefont {A.~J.}\ \bibnamefont {Merz}},\
  and\ \bibinfo {author} {\bibfnamefont {S.~L.}\ \bibnamefont {Keller}},\
  }\bibfield  {title} {\bibinfo {title} {Hallmarks of {Reversible} {Separation}
  of {Living}, {Unperturbed} {Cell} {Membranes} into {Two} {Liquid} {Phases}},\
  }\href {https://doi.org/10.1016/j.bpj.2017.09.029} {\bibfield  {journal}
  {\bibinfo  {journal} {Biophysical Journal}\ }\textbf {\bibinfo {volume}
  {113}},\ \bibinfo {pages} {2425} (\bibinfo {year} {2017})}\BibitemShut
  {NoStop}%
\bibitem [{\citenamefont {Cornell}\ \emph {et~al.}(2018)\citenamefont
  {Cornell}, \citenamefont {Skinkle}, \citenamefont {He}, \citenamefont
  {Levental}, \citenamefont {Levental},\ and\ \citenamefont
  {Keller}}]{cornell_tuning_2018}%
  \BibitemOpen
  \bibfield  {author} {\bibinfo {author} {\bibfnamefont {C.~E.}\ \bibnamefont
  {Cornell}}, \bibinfo {author} {\bibfnamefont {A.~D.}\ \bibnamefont
  {Skinkle}}, \bibinfo {author} {\bibfnamefont {S.}~\bibnamefont {He}},
  \bibinfo {author} {\bibfnamefont {I.}~\bibnamefont {Levental}}, \bibinfo
  {author} {\bibfnamefont {K.~R.}\ \bibnamefont {Levental}},\ and\ \bibinfo
  {author} {\bibfnamefont {S.~L.}\ \bibnamefont {Keller}},\ }\bibfield  {title}
  {\bibinfo {title} {Tuning {Length} {Scales} of {Small} {Domains} in
  {Cell}-{Derived} {Membranes} and {Synthetic} {Model} {Membranes}},\ }\href
  {https://doi.org/10.1016/j.bpj.2018.06.027} {\bibfield  {journal} {\bibinfo
  {journal} {Biophysical Journal}\ }\textbf {\bibinfo {volume} {115}},\
  \bibinfo {pages} {690} (\bibinfo {year} {2018})}\BibitemShut {NoStop}%
\bibitem [{\citenamefont {Banjade}\ and\ \citenamefont
  {Rosen}(2014)}]{banjade_phase_2014}%
  \BibitemOpen
  \bibfield  {author} {\bibinfo {author} {\bibfnamefont {S.}~\bibnamefont
  {Banjade}}\ and\ \bibinfo {author} {\bibfnamefont {M.~K.}\ \bibnamefont
  {Rosen}},\ }\bibfield  {title} {\bibinfo {title} {Phase transitions of
  multivalent proteins can promote clustering of membrane receptors},\ }\href
  {https://doi.org/10.7554/eLife.04123} {\bibfield  {journal} {\bibinfo
  {journal} {eLife}\ }\textbf {\bibinfo {volume} {3}},\ \bibinfo {pages}
  {e04123} (\bibinfo {year} {2014})},\ \bibinfo {note} {publisher: eLife
  Sciences Publications, Ltd}\BibitemShut {NoStop}%
\bibitem [{\citenamefont {Miller}(2018)}]{miller_membrane_2018}%
  \BibitemOpen
  \bibfield  {author} {\bibinfo {author} {\bibfnamefont {J.~L.}\ \bibnamefont
  {Miller}},\ }\bibfield  {title} {\bibinfo {title} {Membrane phase demixing
  seen in living cells},\ }\href {https://doi.org/10.1063/PT.3.3838} {\bibfield
   {journal} {\bibinfo  {journal} {Physics Today}\ }\textbf {\bibinfo {volume}
  {71}},\ \bibinfo {pages} {21} (\bibinfo {year} {2018})}\BibitemShut {NoStop}%
\bibitem [{\citenamefont {Li}\ \emph {et~al.}(2012)\citenamefont {Li},
  \citenamefont {Banjade}, \citenamefont {Cheng}, \citenamefont {Kim},
  \citenamefont {Chen}, \citenamefont {Guo}, \citenamefont {Llaguno},
  \citenamefont {Hollingsworth}, \citenamefont {King}, \citenamefont {Banani},
  \citenamefont {Russo}, \citenamefont {Jiang}, \citenamefont {Nixon},\ and\
  \citenamefont {Rosen}}]{li_phase_2012}%
  \BibitemOpen
  \bibfield  {author} {\bibinfo {author} {\bibfnamefont {P.}~\bibnamefont
  {Li}}, \bibinfo {author} {\bibfnamefont {S.}~\bibnamefont {Banjade}},
  \bibinfo {author} {\bibfnamefont {H.-C.}\ \bibnamefont {Cheng}}, \bibinfo
  {author} {\bibfnamefont {S.}~\bibnamefont {Kim}}, \bibinfo {author}
  {\bibfnamefont {B.}~\bibnamefont {Chen}}, \bibinfo {author} {\bibfnamefont
  {L.}~\bibnamefont {Guo}}, \bibinfo {author} {\bibfnamefont {M.}~\bibnamefont
  {Llaguno}}, \bibinfo {author} {\bibfnamefont {J.~V.}\ \bibnamefont
  {Hollingsworth}}, \bibinfo {author} {\bibfnamefont {D.~S.}\ \bibnamefont
  {King}}, \bibinfo {author} {\bibfnamefont {S.~F.}\ \bibnamefont {Banani}},
  \bibinfo {author} {\bibfnamefont {P.~S.}\ \bibnamefont {Russo}}, \bibinfo
  {author} {\bibfnamefont {Q.-X.}\ \bibnamefont {Jiang}}, \bibinfo {author}
  {\bibfnamefont {B.~T.}\ \bibnamefont {Nixon}},\ and\ \bibinfo {author}
  {\bibfnamefont {M.~K.}\ \bibnamefont {Rosen}},\ }\bibfield  {title} {\bibinfo
  {title} {Phase transitions in the assembly of multivalent signalling
  proteins},\ }\href {https://doi.org/10.1038/nature10879} {\bibfield
  {journal} {\bibinfo  {journal} {Nature}\ }\textbf {\bibinfo {volume} {483}},\
  \bibinfo {pages} {336} (\bibinfo {year} {2012})},\ \bibinfo {note} {number:
  7389 Publisher: Nature Publishing Group}\BibitemShut {NoStop}%
\bibitem [{\citenamefont {Patel}\ \emph {et~al.}(2015)\citenamefont {Patel},
  \citenamefont {Lee}, \citenamefont {Jawerth}, \citenamefont {Maharana},
  \citenamefont {Jahnel}, \citenamefont {Hein}, \citenamefont {Stoynov},
  \citenamefont {Mahamid}, \citenamefont {Saha}, \citenamefont {Franzmann},
  \citenamefont {Pozniakovski}, \citenamefont {Poser}, \citenamefont
  {Maghelli}, \citenamefont {Royer}, \citenamefont {Weigert}, \citenamefont
  {Myers}, \citenamefont {Grill}, \citenamefont {Drechsel}, \citenamefont
  {Hyman},\ and\ \citenamefont {Alberti}}]{patel_liquid--solid_2015}%
  \BibitemOpen
  \bibfield  {author} {\bibinfo {author} {\bibfnamefont {A.}~\bibnamefont
  {Patel}}, \bibinfo {author} {\bibfnamefont {H.~O.}\ \bibnamefont {Lee}},
  \bibinfo {author} {\bibfnamefont {L.}~\bibnamefont {Jawerth}}, \bibinfo
  {author} {\bibfnamefont {S.}~\bibnamefont {Maharana}}, \bibinfo {author}
  {\bibfnamefont {M.}~\bibnamefont {Jahnel}}, \bibinfo {author} {\bibfnamefont
  {M.~Y.}\ \bibnamefont {Hein}}, \bibinfo {author} {\bibfnamefont
  {S.}~\bibnamefont {Stoynov}}, \bibinfo {author} {\bibfnamefont
  {J.}~\bibnamefont {Mahamid}}, \bibinfo {author} {\bibfnamefont
  {S.}~\bibnamefont {Saha}}, \bibinfo {author} {\bibfnamefont {T.~M.}\
  \bibnamefont {Franzmann}}, \bibinfo {author} {\bibfnamefont {A.}~\bibnamefont
  {Pozniakovski}}, \bibinfo {author} {\bibfnamefont {I.}~\bibnamefont {Poser}},
  \bibinfo {author} {\bibfnamefont {N.}~\bibnamefont {Maghelli}}, \bibinfo
  {author} {\bibfnamefont {L.~A.}\ \bibnamefont {Royer}}, \bibinfo {author}
  {\bibfnamefont {M.}~\bibnamefont {Weigert}}, \bibinfo {author} {\bibfnamefont
  {E.~W.}\ \bibnamefont {Myers}}, \bibinfo {author} {\bibfnamefont
  {S.}~\bibnamefont {Grill}}, \bibinfo {author} {\bibfnamefont
  {D.}~\bibnamefont {Drechsel}}, \bibinfo {author} {\bibfnamefont {A.~A.}\
  \bibnamefont {Hyman}},\ and\ \bibinfo {author} {\bibfnamefont
  {S.}~\bibnamefont {Alberti}},\ }\bibfield  {title} {\bibinfo {title} {A
  {Liquid}-to-{Solid} {Phase} {Transition} of the {ALS} {Protein} {FUS}
  {Accelerated} by {Disease} {Mutation}},\ }\href
  {https://doi.org/10.1016/j.cell.2015.07.047} {\bibfield  {journal} {\bibinfo
  {journal} {Cell}\ }\textbf {\bibinfo {volume} {162}},\ \bibinfo {pages}
  {1066} (\bibinfo {year} {2015})},\ \bibinfo {note} {publisher:
  Elsevier}\BibitemShut {NoStop}%
\bibitem [{\citenamefont {Bieler}\ \emph {et~al.}(2011)\citenamefont {Bieler},
  \citenamefont {Pozzorini},\ and\ \citenamefont
  {Naef}}]{bieler_whole-embryo_2011}%
  \BibitemOpen
  \bibfield  {author} {\bibinfo {author} {\bibfnamefont {J.}~\bibnamefont
  {Bieler}}, \bibinfo {author} {\bibfnamefont {C.}~\bibnamefont {Pozzorini}},\
  and\ \bibinfo {author} {\bibfnamefont {F.}~\bibnamefont {Naef}},\ }\bibfield
  {title} {\bibinfo {title} {Whole-{Embryo} {Modeling} of {Early}
  {Segmentation} in {Drosophila} {Identifies} {Robust} and {Fragile}
  {Expression} {Domains}},\ }\href {https://doi.org/10.1016/j.bpj.2011.05.060}
  {\bibfield  {journal} {\bibinfo  {journal} {Biophys J}\ }\textbf {\bibinfo
  {volume} {101}},\ \bibinfo {pages} {287} (\bibinfo {year}
  {2011})}\BibitemShut {NoStop}%
\bibitem [{\citenamefont {{Hyman Anthony A}}\ and\ \citenamefont
  {Wheeler}()}]{hyman_anthony_a_controlling_nodate}%
  \BibitemOpen
  \bibfield  {author} {\bibinfo {author} {\bibnamefont {{Hyman Anthony A}}}\
  and\ \bibinfo {author} {\bibfnamefont {R.}~\bibnamefont {Wheeler}},\ }\href
  {https://royalsocietypublishing.org/doi/10.1098/rstb.2017.0193} {\bibinfo
  {title} {Controlling compartmentalization by non-membrane-bound organelles
  {\textbar} {Philosophical} {Transactions} of the {Royal} {Society} {B}:
  {Biological} {Sciences}}}\BibitemShut {NoStop}%
\bibitem [{\citenamefont {Su}\ \emph {et~al.}(2016)\citenamefont {Su},
  \citenamefont {Ditlev}, \citenamefont {Hui}, \citenamefont {Xing},
  \citenamefont {Banjade}, \citenamefont {Okrut}, \citenamefont {King},
  \citenamefont {Taunton}, \citenamefont {Rosen},\ and\ \citenamefont
  {Vale}}]{su_phase_2016}%
  \BibitemOpen
  \bibfield  {author} {\bibinfo {author} {\bibfnamefont {X.}~\bibnamefont
  {Su}}, \bibinfo {author} {\bibfnamefont {J.~A.}\ \bibnamefont {Ditlev}},
  \bibinfo {author} {\bibfnamefont {E.}~\bibnamefont {Hui}}, \bibinfo {author}
  {\bibfnamefont {W.}~\bibnamefont {Xing}}, \bibinfo {author} {\bibfnamefont
  {S.}~\bibnamefont {Banjade}}, \bibinfo {author} {\bibfnamefont
  {J.}~\bibnamefont {Okrut}}, \bibinfo {author} {\bibfnamefont {D.~S.}\
  \bibnamefont {King}}, \bibinfo {author} {\bibfnamefont {J.}~\bibnamefont
  {Taunton}}, \bibinfo {author} {\bibfnamefont {M.~K.}\ \bibnamefont {Rosen}},\
  and\ \bibinfo {author} {\bibfnamefont {R.~D.}\ \bibnamefont {Vale}},\
  }\bibfield  {title} {\bibinfo {title} {Phase separation of signaling
  molecules promotes {T} cell receptor signal transduction},\ }\href
  {https://doi.org/10.1126/science.aad9964} {\bibfield  {journal} {\bibinfo
  {journal} {Science}\ }\textbf {\bibinfo {volume} {352}},\ \bibinfo {pages}
  {595} (\bibinfo {year} {2016})},\ \bibinfo {note} {publisher: American
  Association for the Advancement of Science}\BibitemShut {NoStop}%
\bibitem [{\citenamefont {Brangwynne}\ \emph {et~al.}(2009)\citenamefont
  {Brangwynne}, \citenamefont {Eckmann}, \citenamefont {Courson}, \citenamefont
  {Rybarska}, \citenamefont {Hoege}, \citenamefont {Gharakhani}, \citenamefont
  {Jülicher},\ and\ \citenamefont {Hyman}}]{brangwynne_germline_2009}%
  \BibitemOpen
  \bibfield  {author} {\bibinfo {author} {\bibfnamefont {C.~P.}\ \bibnamefont
  {Brangwynne}}, \bibinfo {author} {\bibfnamefont {C.~R.}\ \bibnamefont
  {Eckmann}}, \bibinfo {author} {\bibfnamefont {D.~S.}\ \bibnamefont
  {Courson}}, \bibinfo {author} {\bibfnamefont {A.}~\bibnamefont {Rybarska}},
  \bibinfo {author} {\bibfnamefont {C.}~\bibnamefont {Hoege}}, \bibinfo
  {author} {\bibfnamefont {J.}~\bibnamefont {Gharakhani}}, \bibinfo {author}
  {\bibfnamefont {F.}~\bibnamefont {Jülicher}},\ and\ \bibinfo {author}
  {\bibfnamefont {A.~A.}\ \bibnamefont {Hyman}},\ }\bibfield  {title} {\bibinfo
  {title} {Germline {P} {Granules} {Are} {Liquid} {Droplets} {That} {Localize}
  by {Controlled} {Dissolution}/{Condensation}},\ }\href
  {https://doi.org/10.1126/science.1172046} {\bibfield  {journal} {\bibinfo
  {journal} {Science}\ }\textbf {\bibinfo {volume} {324}},\ \bibinfo {pages}
  {1729} (\bibinfo {year} {2009})},\ \bibinfo {note} {publisher: American
  Association for the Advancement of Science}\BibitemShut {NoStop}%
\bibitem [{\citenamefont {Brangwynne}\ \emph {et~al.}(2011)\citenamefont
  {Brangwynne}, \citenamefont {Mitchison},\ and\ \citenamefont
  {Hyman}}]{hyman_active_2011}%
  \BibitemOpen
  \bibfield  {author} {\bibinfo {author} {\bibfnamefont {C.~P.}\ \bibnamefont
  {Brangwynne}}, \bibinfo {author} {\bibfnamefont {T.~J.}\ \bibnamefont
  {Mitchison}},\ and\ \bibinfo {author} {\bibfnamefont {A.~A.}\ \bibnamefont
  {Hyman}},\ }\bibfield  {title} {\bibinfo {title} {Active liquid-like behavior
  of nucleoli determines their size and shape in <i>xenopus laevis</i>
  oocytes},\ }\href {https://doi.org/10.1073/pnas.1017150108} {\bibfield
  {journal} {\bibinfo  {journal} {Proceedings of the National Academy of
  Sciences}\ }\textbf {\bibinfo {volume} {108}},\ \bibinfo {pages} {4334}
  (\bibinfo {year} {2011})},\ \Eprint
  {https://arxiv.org/abs/https://www.pnas.org/doi/pdf/10.1073/pnas.1017150108}
  {https://www.pnas.org/doi/pdf/10.1073/pnas.1017150108} \BibitemShut {NoStop}%
\bibitem [{\citenamefont {Cates}\ and\ \citenamefont
  {Tailleur}(2015)}]{cates_motility-induced_2015}%
  \BibitemOpen
  \bibfield  {author} {\bibinfo {author} {\bibfnamefont {M.~E.}\ \bibnamefont
  {Cates}}\ and\ \bibinfo {author} {\bibfnamefont {J.}~\bibnamefont
  {Tailleur}},\ }\bibfield  {title} {\bibinfo {title} {Motility-{Induced}
  {Phase} {Separation}},\ }\href
  {https://doi.org/10.1146/annurev-conmatphys-031214-014710} {\bibfield
  {journal} {\bibinfo  {journal} {Annual Review of Condensed Matter Physics}\
  }\textbf {\bibinfo {volume} {6}},\ \bibinfo {pages} {219} (\bibinfo {year}
  {2015})},\ \bibinfo {note} {\_eprint:
  https://doi.org/10.1146/annurev-conmatphys-031214-014710}\BibitemShut
  {NoStop}%
\bibitem [{\citenamefont {Marchetti}\ \emph {et~al.}(2013)\citenamefont
  {Marchetti}, \citenamefont {Joanny}, \citenamefont {Ramaswamy}, \citenamefont
  {Liverpool}, \citenamefont {Prost}, \citenamefont {Rao},\ and\ \citenamefont
  {Simha}}]{marchetti_hydrodynamics_2013}%
  \BibitemOpen
  \bibfield  {author} {\bibinfo {author} {\bibfnamefont {M.~C.}\ \bibnamefont
  {Marchetti}}, \bibinfo {author} {\bibfnamefont {J.~F.}\ \bibnamefont
  {Joanny}}, \bibinfo {author} {\bibfnamefont {S.}~\bibnamefont {Ramaswamy}},
  \bibinfo {author} {\bibfnamefont {T.~B.}\ \bibnamefont {Liverpool}}, \bibinfo
  {author} {\bibfnamefont {J.}~\bibnamefont {Prost}}, \bibinfo {author}
  {\bibfnamefont {M.}~\bibnamefont {Rao}},\ and\ \bibinfo {author}
  {\bibfnamefont {R.~A.}\ \bibnamefont {Simha}},\ }\bibfield  {title} {\bibinfo
  {title} {Hydrodynamics of soft active matter},\ }\href
  {https://doi.org/10.1103/RevModPhys.85.1143} {\bibfield  {journal} {\bibinfo
  {journal} {Rev. Mod. Phys.}\ }\textbf {\bibinfo {volume} {85}},\ \bibinfo
  {pages} {1143} (\bibinfo {year} {2013})},\ \bibinfo {note} {publisher:
  American Physical Society}\BibitemShut {NoStop}%
\bibitem [{\citenamefont {Yang}\ \emph {et~al.}(2014)\citenamefont {Yang},
  \citenamefont {Manning},\ and\ \citenamefont
  {Marchetti}}]{yang2014aggregation}%
  \BibitemOpen
  \bibfield  {author} {\bibinfo {author} {\bibfnamefont {X.}~\bibnamefont
  {Yang}}, \bibinfo {author} {\bibfnamefont {M.~L.}\ \bibnamefont {Manning}},\
  and\ \bibinfo {author} {\bibfnamefont {M.~C.}\ \bibnamefont {Marchetti}},\
  }\bibfield  {title} {\bibinfo {title} {Aggregation and segregation of
  confined active particles},\ }\href@noop {} {\bibfield  {journal} {\bibinfo
  {journal} {Soft matter}\ }\textbf {\bibinfo {volume} {10}},\ \bibinfo {pages}
  {6477} (\bibinfo {year} {2014})}\BibitemShut {NoStop}%
\bibitem [{\citenamefont {V}\ \emph {et~al.}(2021)\citenamefont {V},
  \citenamefont {Lin},\ and\ \citenamefont {Maiti}}]{nayana_soft_2021}%
  \BibitemOpen
  \bibfield  {author} {\bibinfo {author} {\bibfnamefont {N.}~\bibnamefont {V}},
  \bibinfo {author} {\bibfnamefont {S.-T.}\ \bibnamefont {Lin}},\ and\ \bibinfo
  {author} {\bibfnamefont {P.~K.}\ \bibnamefont {Maiti}},\ }\href
  {https://doi.org/10.48550/ARXIV.2109.00415} {\bibinfo {title} {Phase behavior
  of active and passive dumbbells}} (\bibinfo {year} {2021})\BibitemShut
  {NoStop}%
\bibitem [{\citenamefont {Ilker}\ and\ \citenamefont
  {Joanny}(2020)}]{ilker_2020}%
  \BibitemOpen
  \bibfield  {author} {\bibinfo {author} {\bibfnamefont {E.}~\bibnamefont
  {Ilker}}\ and\ \bibinfo {author} {\bibfnamefont {J.-F.}\ \bibnamefont
  {Joanny}},\ }\bibfield  {title} {\bibinfo {title} {Phase separation and
  nucleation in mixtures of particles with different temperatures},\ }\bibfield
   {journal} {\bibinfo  {journal} {Physical Review Research}\ }\textbf
  {\bibinfo {volume} {2}},\ \href
  {https://doi.org/10.1103/physrevresearch.2.023200}
  {10.1103/physrevresearch.2.023200} (\bibinfo {year} {2020})\BibitemShut
  {NoStop}%
\bibitem [{\citenamefont {Xu}\ \emph {et~al.}(2019)\citenamefont {Xu},
  \citenamefont {Ross}, \citenamefont {Valdez},\ and\ \citenamefont
  {Sen}}]{xu_direct_2019}%
  \BibitemOpen
  \bibfield  {author} {\bibinfo {author} {\bibfnamefont {M.}~\bibnamefont
  {Xu}}, \bibinfo {author} {\bibfnamefont {J.~L.}\ \bibnamefont {Ross}},
  \bibinfo {author} {\bibfnamefont {L.}~\bibnamefont {Valdez}},\ and\ \bibinfo
  {author} {\bibfnamefont {A.}~\bibnamefont {Sen}},\ }\bibfield  {title}
  {\bibinfo {title} {Direct {Single} {Molecule} {Imaging} of {Enhanced}
  {Enzyme} {Diffusion}},\ }\href
  {https://doi.org/10.1103/PhysRevLett.123.128101} {\bibfield  {journal}
  {\bibinfo  {journal} {Phys. Rev. Lett.}\ }\textbf {\bibinfo {volume} {123}},\
  \bibinfo {pages} {128101} (\bibinfo {year} {2019})},\ \bibinfo {note}
  {publisher: American Physical Society}\BibitemShut {NoStop}%
\bibitem [{\citenamefont {Gartner}(2022)}]{ZevGartner}%
  \BibitemOpen
  \bibfield  {author} {\bibinfo {author} {\bibfnamefont {Z.}~\bibnamefont
  {Gartner}},\ }\href@noop {} {}\bibinfo {howpublished} {private communication}
  (\bibinfo {year} {2022})\BibitemShut {NoStop}%
\bibitem [{\citenamefont {Henkes}\ \emph {et~al.}(2011)\citenamefont {Henkes},
  \citenamefont {Fily},\ and\ \citenamefont {Marchetti}}]{henkes_2011}%
  \BibitemOpen
  \bibfield  {author} {\bibinfo {author} {\bibfnamefont {S.}~\bibnamefont
  {Henkes}}, \bibinfo {author} {\bibfnamefont {Y.}~\bibnamefont {Fily}},\ and\
  \bibinfo {author} {\bibfnamefont {M.~C.}\ \bibnamefont {Marchetti}},\
  }\bibfield  {title} {\bibinfo {title} {Active jamming: Self-propelled soft
  particles at high density},\ }\bibfield  {journal} {\bibinfo  {journal}
  {Physical Review E}\ }\textbf {\bibinfo {volume} {84}},\ \href
  {https://doi.org/10.1103/physreve.84.040301} {10.1103/physreve.84.040301}
  (\bibinfo {year} {2011})\BibitemShut {NoStop}%
\bibitem [{\citenamefont {Morse}\ \emph {et~al.}(2021)\citenamefont {Morse},
  \citenamefont {Roy}, \citenamefont {Agoritsas}, \citenamefont {Stanifer},
  \citenamefont {Corwin},\ and\ \citenamefont {Manning}}]{morse_direct_2021}%
  \BibitemOpen
  \bibfield  {author} {\bibinfo {author} {\bibfnamefont {P.~K.}\ \bibnamefont
  {Morse}}, \bibinfo {author} {\bibfnamefont {S.}~\bibnamefont {Roy}}, \bibinfo
  {author} {\bibfnamefont {E.}~\bibnamefont {Agoritsas}}, \bibinfo {author}
  {\bibfnamefont {E.}~\bibnamefont {Stanifer}}, \bibinfo {author}
  {\bibfnamefont {E.~I.}\ \bibnamefont {Corwin}},\ and\ \bibinfo {author}
  {\bibfnamefont {M.~L.}\ \bibnamefont {Manning}},\ }\bibfield  {title}
  {\bibinfo {title} {A direct link between active matter and sheared granular
  systems},\ }\href {https://doi.org/10.1073/pnas.2019909118} {\bibfield
  {journal} {\bibinfo  {journal} {Proceedings of the National Academy of
  Sciences}\ }\textbf {\bibinfo {volume} {118}},\ \bibinfo {pages}
  {e2019909118} (\bibinfo {year} {2021})},\ \bibinfo {note} {publisher:
  Proceedings of the National Academy of Sciences}\BibitemShut {NoStop}%
\bibitem [{\citenamefont {Mandal}\ \emph {et~al.}(2020)\citenamefont {Mandal},
  \citenamefont {Bhuyan}, \citenamefont {Chaudhuri}, \citenamefont {Dasgupta},\
  and\ \citenamefont {Rao}}]{mandal_extreme_2020}%
  \BibitemOpen
  \bibfield  {author} {\bibinfo {author} {\bibfnamefont {R.}~\bibnamefont
  {Mandal}}, \bibinfo {author} {\bibfnamefont {P.~J.}\ \bibnamefont {Bhuyan}},
  \bibinfo {author} {\bibfnamefont {P.}~\bibnamefont {Chaudhuri}}, \bibinfo
  {author} {\bibfnamefont {C.}~\bibnamefont {Dasgupta}},\ and\ \bibinfo
  {author} {\bibfnamefont {M.}~\bibnamefont {Rao}},\ }\bibfield  {title}
  {\bibinfo {title} {Extreme active matter at high densities},\ }\href
  {https://doi.org/10.1038/s41467-020-16130-x} {\bibfield  {journal} {\bibinfo
  {journal} {Nature Communications}\ }\textbf {\bibinfo {volume} {11}},\
  \bibinfo {pages} {2581} (\bibinfo {year} {2020})},\ \bibinfo {note} {number:
  1 Publisher: Nature Publishing Group}\BibitemShut {NoStop}%
\bibitem [{\citenamefont {Bi}\ \emph {et~al.}(2016)\citenamefont {Bi},
  \citenamefont {Yang}, \citenamefont {Marchetti},\ and\ \citenamefont
  {Manning}}]{bi_motility-driven_2016}%
  \BibitemOpen
  \bibfield  {author} {\bibinfo {author} {\bibfnamefont {D.}~\bibnamefont
  {Bi}}, \bibinfo {author} {\bibfnamefont {X.}~\bibnamefont {Yang}}, \bibinfo
  {author} {\bibfnamefont {M.~C.}\ \bibnamefont {Marchetti}},\ and\ \bibinfo
  {author} {\bibfnamefont {M.~L.}\ \bibnamefont {Manning}},\ }\bibfield
  {title} {\bibinfo {title} {Motility-{Driven} {Glass} and {Jamming}
  {Transitions} in {Biological} {Tissues}},\ }\href
  {https://doi.org/10.1103/PhysRevX.6.021011} {\bibfield  {journal} {\bibinfo
  {journal} {Physical Review X}\ }\textbf {\bibinfo {volume} {6}},\ \bibinfo
  {pages} {021011} (\bibinfo {year} {2016})}\BibitemShut {NoStop}%
\bibitem [{\citenamefont {Sahu}\ \emph {et~al.}(2020)\citenamefont {Sahu},
  \citenamefont {Sussman}, \citenamefont {Rübsam}, \citenamefont {Mertz},
  \citenamefont {Horsley}, \citenamefont {Dufresne}, \citenamefont {Niessen},
  \citenamefont {Marchetti}, \citenamefont {Manning},\ and\ \citenamefont
  {Schwarz}}]{sahu_small-scale_2020}%
  \BibitemOpen
  \bibfield  {author} {\bibinfo {author} {\bibfnamefont {P.}~\bibnamefont
  {Sahu}}, \bibinfo {author} {\bibfnamefont {D.~M.}\ \bibnamefont {Sussman}},
  \bibinfo {author} {\bibfnamefont {M.}~\bibnamefont {Rübsam}}, \bibinfo
  {author} {\bibfnamefont {A.~F.}\ \bibnamefont {Mertz}}, \bibinfo {author}
  {\bibfnamefont {V.}~\bibnamefont {Horsley}}, \bibinfo {author} {\bibfnamefont
  {E.~R.}\ \bibnamefont {Dufresne}}, \bibinfo {author} {\bibfnamefont {C.~M.}\
  \bibnamefont {Niessen}}, \bibinfo {author} {\bibfnamefont {M.~C.}\
  \bibnamefont {Marchetti}}, \bibinfo {author} {\bibfnamefont {M.~L.}\
  \bibnamefont {Manning}},\ and\ \bibinfo {author} {\bibfnamefont {J.~M.}\
  \bibnamefont {Schwarz}},\ }\bibfield  {title} {\bibinfo {title} {Small-scale
  demixing in confluent biological tissues},\ }\href
  {https://doi.org/10.1039/C9SM01084J} {\bibfield  {journal} {\bibinfo
  {journal} {Soft Matter}\ }\textbf {\bibinfo {volume} {16}},\ \bibinfo {pages}
  {3325} (\bibinfo {year} {2020})}\BibitemShut {NoStop}%
\bibitem [{\citenamefont {Smrek}\ and\ \citenamefont
  {Kremer}(2017)}]{smrek_small_2017}%
  \BibitemOpen
  \bibfield  {author} {\bibinfo {author} {\bibfnamefont {J.}~\bibnamefont
  {Smrek}}\ and\ \bibinfo {author} {\bibfnamefont {K.}~\bibnamefont {Kremer}},\
  }\bibfield  {title} {\bibinfo {title} {Small {Activity} {Differences} {Drive}
  {Phase} {Separation} in {Active}-{Passive} {Polymer} {Mixtures}},\ }\href
  {https://doi.org/10.1103/PhysRevLett.118.098002} {\bibfield  {journal}
  {\bibinfo  {journal} {Physical Review Letters}\ }\textbf {\bibinfo {volume}
  {118}},\ \bibinfo {pages} {098002} (\bibinfo {year} {2017})},\ \bibinfo
  {note} {publisher: American Physical Society}\BibitemShut {NoStop}%
\bibitem [{\citenamefont {Petridou}\ \emph {et~al.}(2021)\citenamefont
  {Petridou}, \citenamefont {Corominas-Murtra}, \citenamefont {Heisenberg},\
  and\ \citenamefont {Hannezo}}]{petridou2021rigidity}%
  \BibitemOpen
  \bibfield  {author} {\bibinfo {author} {\bibfnamefont {N.~I.}\ \bibnamefont
  {Petridou}}, \bibinfo {author} {\bibfnamefont {B.}~\bibnamefont
  {Corominas-Murtra}}, \bibinfo {author} {\bibfnamefont {C.-P.}\ \bibnamefont
  {Heisenberg}},\ and\ \bibinfo {author} {\bibfnamefont {E.}~\bibnamefont
  {Hannezo}},\ }\bibfield  {title} {\bibinfo {title} {Rigidity percolation
  uncovers a structural basis for embryonic tissue phase transitions},\
  }\href@noop {} {\bibfield  {journal} {\bibinfo  {journal} {Cell}\ }\textbf
  {\bibinfo {volume} {184}},\ \bibinfo {pages} {1914} (\bibinfo {year}
  {2021})}\BibitemShut {NoStop}%
\bibitem [{\citenamefont {Ranft}\ \emph {et~al.}(2010)\citenamefont {Ranft},
  \citenamefont {Basan}, \citenamefont {Elgeti}, \citenamefont {Joanny},
  \citenamefont {Prost},\ and\ \citenamefont
  {J{\"u}licher}}]{ranft2010fluidization}%
  \BibitemOpen
  \bibfield  {author} {\bibinfo {author} {\bibfnamefont {J.}~\bibnamefont
  {Ranft}}, \bibinfo {author} {\bibfnamefont {M.}~\bibnamefont {Basan}},
  \bibinfo {author} {\bibfnamefont {J.}~\bibnamefont {Elgeti}}, \bibinfo
  {author} {\bibfnamefont {J.-F.}\ \bibnamefont {Joanny}}, \bibinfo {author}
  {\bibfnamefont {J.}~\bibnamefont {Prost}},\ and\ \bibinfo {author}
  {\bibfnamefont {F.}~\bibnamefont {J{\"u}licher}},\ }\bibfield  {title}
  {\bibinfo {title} {Fluidization of tissues by cell division and apoptosis},\
  }\href@noop {} {\bibfield  {journal} {\bibinfo  {journal} {Proceedings of the
  National Academy of Sciences}\ }\textbf {\bibinfo {volume} {107}},\ \bibinfo
  {pages} {20863} (\bibinfo {year} {2010})}\BibitemShut {NoStop}%
\bibitem [{\citenamefont {Ellenbroek}(2013)}]{ellenbroek2013unusual}%
  \BibitemOpen
  \bibfield  {author} {\bibinfo {author} {\bibfnamefont {W.~G.}\ \bibnamefont
  {Ellenbroek}},\ }\bibfield  {title} {\bibinfo {title} {Unusual order in
  squeezed repulsive spheres},\ }in\ \href@noop {} {\emph {\bibinfo {booktitle}
  {APS March Meeting Abstracts}}},\ Vol.\ \bibinfo {volume} {2013}\ (\bibinfo
  {year} {2013})\ pp.\ \bibinfo {pages} {W29--010}\BibitemShut {NoStop}%
\bibitem [{\citenamefont {Takatori}\ and\ \citenamefont
  {Brady}(2015)}]{takatori_towards_2015}%
  \BibitemOpen
  \bibfield  {author} {\bibinfo {author} {\bibfnamefont {S.~C.}\ \bibnamefont
  {Takatori}}\ and\ \bibinfo {author} {\bibfnamefont {J.~F.}\ \bibnamefont
  {Brady}},\ }\bibfield  {title} {\bibinfo {title} {Towards a thermodynamics of
  active matter},\ }\href {https://doi.org/10.1103/PhysRevE.91.032117}
  {\bibfield  {journal} {\bibinfo  {journal} {Physical Review E}\ }\textbf
  {\bibinfo {volume} {91}},\ \bibinfo {pages} {032117} (\bibinfo {year}
  {2015})}\BibitemShut {NoStop}%
\bibitem [{\citenamefont {Canty}\ \emph {et~al.}(2017)\citenamefont {Canty},
  \citenamefont {Zarour}, \citenamefont {Kashkooli}, \citenamefont
  {François},\ and\ \citenamefont {Fagotto}}]{canty_sorting_2017}%
  \BibitemOpen
  \bibfield  {author} {\bibinfo {author} {\bibfnamefont {L.}~\bibnamefont
  {Canty}}, \bibinfo {author} {\bibfnamefont {E.}~\bibnamefont {Zarour}},
  \bibinfo {author} {\bibfnamefont {L.}~\bibnamefont {Kashkooli}}, \bibinfo
  {author} {\bibfnamefont {P.}~\bibnamefont {François}},\ and\ \bibinfo
  {author} {\bibfnamefont {F.}~\bibnamefont {Fagotto}},\ }\bibfield  {title}
  {\bibinfo {title} {Sorting at embryonic boundaries requires high heterotypic
  interfacial tension},\ }\href {https://doi.org/10.1038/s41467-017-00146-x}
  {\bibfield  {journal} {\bibinfo  {journal} {Nat Commun}\ }\textbf {\bibinfo
  {volume} {8}},\ \bibinfo {pages} {157} (\bibinfo {year} {2017})},\ \bibinfo
  {note} {number: 1 Publisher: Nature Publishing Group}\BibitemShut {NoStop}%
\bibitem [{\citenamefont {Sussman}\ \emph {et~al.}(2018)\citenamefont
  {Sussman}, \citenamefont {Schwarz}, \citenamefont {Marchetti},\ and\
  \citenamefont {Manning}}]{sussman_soft_2018}%
  \BibitemOpen
  \bibfield  {author} {\bibinfo {author} {\bibfnamefont {D.~M.}\ \bibnamefont
  {Sussman}}, \bibinfo {author} {\bibfnamefont {J.}~\bibnamefont {Schwarz}},
  \bibinfo {author} {\bibfnamefont {M.~C.}\ \bibnamefont {Marchetti}},\ and\
  \bibinfo {author} {\bibfnamefont {M.~L.}\ \bibnamefont {Manning}},\
  }\bibfield  {title} {\bibinfo {title} {Soft yet {Sharp} {Interfaces} in a
  {Vertex} {Model} of {Confluent} {Tissue}},\ }\href
  {https://doi.org/10.1103/PhysRevLett.120.058001} {\bibfield  {journal}
  {\bibinfo  {journal} {Phys. Rev. Lett.}\ }\textbf {\bibinfo {volume} {120}},\
  \bibinfo {pages} {058001} (\bibinfo {year} {2018})},\ \bibinfo {note}
  {publisher: American Physical Society}\BibitemShut {NoStop}%
\bibitem [{\citenamefont {Dix}\ and\ \citenamefont
  {Verkman}(2008)}]{dix_crowding_2008}%
  \BibitemOpen
  \bibfield  {author} {\bibinfo {author} {\bibfnamefont {J.~A.}\ \bibnamefont
  {Dix}}\ and\ \bibinfo {author} {\bibfnamefont {A.~S.}\ \bibnamefont
  {Verkman}},\ }\bibfield  {title} {\bibinfo {title} {Crowding effects on
  diffusion in solutions and cells},\ }\href
  {https://doi.org/10.1146/annurev.biophys.37.032807.125824} {\bibfield
  {journal} {\bibinfo  {journal} {Annual Review of Biophysics}\ }\textbf
  {\bibinfo {volume} {37}},\ \bibinfo {pages} {247} (\bibinfo {year}
  {2008})}\BibitemShut {NoStop}%
\bibitem [{\citenamefont {André}\ and\ \citenamefont
  {Spruijt}(2020)}]{andre_liquidliquid_2020}%
  \BibitemOpen
  \bibfield  {author} {\bibinfo {author} {\bibfnamefont {A.~A.~M.}\
  \bibnamefont {André}}\ and\ \bibinfo {author} {\bibfnamefont
  {E.}~\bibnamefont {Spruijt}},\ }\bibfield  {title} {\bibinfo {title}
  {Liquid–{Liquid} {Phase} {Separation} in {Crowded} {Environments}},\ }\href
  {https://doi.org/10.3390/ijms21165908} {\bibfield  {journal} {\bibinfo
  {journal} {International Journal of Molecular Sciences}\ }\textbf {\bibinfo
  {volume} {21}},\ \bibinfo {pages} {5908} (\bibinfo {year} {2020})},\ \bibinfo
  {note} {number: 16 Publisher: Multidisciplinary Digital Publishing
  Institute}\BibitemShut {NoStop}%
\bibitem [{\citenamefont {Peeples}\ and\ \citenamefont
  {Rosen}(2021)}]{peeples_mechanistic_2021}%
  \BibitemOpen
  \bibfield  {author} {\bibinfo {author} {\bibfnamefont {W.}~\bibnamefont
  {Peeples}}\ and\ \bibinfo {author} {\bibfnamefont {M.~K.}\ \bibnamefont
  {Rosen}},\ }\bibfield  {title} {\bibinfo {title} {Mechanistic dissection of
  increased enzymatic rate in a phase-separated compartment},\ }\href
  {https://doi.org/10.1038/s41589-021-00801-x} {\bibfield  {journal} {\bibinfo
  {journal} {Nature Chemical Biology}\ }\textbf {\bibinfo {volume} {17}},\
  \bibinfo {pages} {693} (\bibinfo {year} {2021})},\ \bibinfo {note} {number: 6
  Publisher: Nature Publishing Group}\BibitemShut {NoStop}%
\end{thebibliography}%

\end{document}